\documentclass[12pt]{article}

\usepackage{newtxtext,newtxmath}

\usepackage{graphicx}
\usepackage[dvipsnames]{xcolor}

\usepackage[letterpaper,margin=1in]{geometry}

\renewenvironment{abstract}
	{\quotation}
	{\endquotation}

\date{}

\makeatletter
\renewcommand{\fnum@figure}{\textbf{Fig. \thefigure}}
\renewcommand{\fnum@table}{\textbf{Table \thetable}}
\makeatother

\usepackage[numbers,super,sort&compress]{natbib}

\usepackage{url}

\def\scititle{
	Observation of angular momentum transfer among crystal lattice modes
}
\title{\bfseries \boldmath \scititle}

\author{
	Olga~Minakova$^{1}$,
    Carolina~Paiva$^{2}$,
    Maximilian~Frenzel$^{1}$,
    Michael~S.~Spencer$^{1}$,\and
    Joanna~M.~Urban$^{1}$,
    Christoph~Ringkamp$^{3,4}$,
    Martin~Wolf$^{1}$,
    Gregor~Mussler$^{3,4}$,\and
    Dominik~M.~Juraschek$^{2,5}$,
    Sebastian~F.~Maehrlein$^{1,6,7\ast}$\and
	\small$^{1}$Department of Physical Chemistry, Fritz Haber Institute of the Max Planck Society, Berlin, 14195, Germany.\and
    \small$^{2}$School of Physics and Astronomy, Tel Aviv University, Tel Aviv, 69978, Israel. \and
    \small$^{3}$Peter Grünberg Institut (PGI-9), Forschungszentrum Jülich, Jülich, 52425, Germany.\and
    \small$^{4}$JARA-Fundamentals of Future Information Technology, Jülich-Aachen Research Alliance,\and
    \small Forschungszentrum Jülich GmbH and RWTH Aachen University, Jülich, 52425, Germany.\and
    \small$^{5}$Department of Applied Physics and Science Education, Eindhoven University of Technology, Eindhoven,\and
    \small 5612 AP, Netherlands. \and
    \small$^{6}$Institute of Radiation Physics, Helmholtz-Zentrum Dresden-Rossendorf, Dresden, 01328, Germany. \and
    \small$^{7}$Institute of Applied Physics, Dresden University of Technology, Dresden, 01062, Germany. \and
	\small$^\ast$Corresponding author. Email: s.maehrlein@hzdr.de\and 
}

\begin{document} 

\maketitle

\newpage
\begin{abstract} \bfseries \boldmath
Transfer of energy and linear momentum between lattice vibrations via anharmonic coupling is an important concept in solid-state physics. However, it remains difficult to directly observe how \textit{angular} momentum is exchanged and conserved among lattice modes, even though these processes are thought to play an important role in achieving magnetization equilibrium and in spin relaxation effects like the Einstein–de Haas effect. Here, we demonstrate and coherently control angular momentum transfer between two lattice modes using the inverse process of anharmonic decay. The observed rotational phonon-phonon Umklapp scattering enforces the conservation of quantized crystal angular momentum, as dictated by the crystal’s discrete rotational symmetry. We thereby experimentally confirm the fundamental analogy between linear and angular momentum conservation in solids. Moreover, we establish axial nonlinear phononics towards ultrafast control of material properties. 
\end{abstract}

\noindent
Fundamental physics is dictated by symmetry principles, linking conservation laws to time, space translation, and rotational invariance. In crystalline solids, discrete translational symmetry enforces the conservation of pseudo linear momentum, also called crystal momentum, forming the basis of Bloch’s theorem, band structure theory, and modern semiconductor-based technology. Therefore, quantized vibrations of the crystalline lattice (phonons) must conserve energy and linear crystal momentum during their interactions, which are only permitted by anharmonic phonon-phonon scattering requiring at least a three-phonon interaction. Such energy and momentum relaxation through anharmonic phonon-phonon interaction is fundamentally required to establish equilibrium states of condensed matter. The observation of phonon Umklapp scattering processes, e.g. governing thermal conductivity, serves as a clear signature of conserved linear crystal momentum imposed by the periodicity of the lattice~\cite{Kittel2005}.

However, the conservation and redistribution of angular momentum within the lattice, fundamentally required to reach magnetization equilibrium~\cite{Einstein1915, Dornes2019, Tauchert2022} and imperative for spin relaxation phenomena in solids, remains a postulate~\cite{Zhang2014, Garanin2015}. Since the discovery of the Einstein-de Haas effect~\cite{Einstein1915}, it has been recognized that electron spin angular momentum must eventually transfer to the lattice as a rigid-body rotation, bridging quantum spin dynamics with classical mechanical angular momentum~\cite{Dornes2019, Zhang2025}. Yet, how this transfer proceeds after the initial spin-lattice coupling\cite{Tauchert2022, Choi2024, Shokeen2024, Mrudul2025} and how it relates to the discrete rotational invariance of the lattice is still a black box to date. In contrast to the continuous rotational invariance of free space, in solids a discrete $n$-fold rotational invariance $C_n$ can only conserve angular momentum modulo $n$ in the form of “pseudo” – here also called “crystal” – angular momentum. Yet, direct experimental observation of phonon trajectories supporting this conceptual analogy between linear and angular momentum conservation in solids has remained elusive. Recently, lattice angular momentum in the form of circularly polarized chiral~\cite{Ishito2022, Ueda2023, Zhu2018} or axial~\cite{Juraschek2025_review} phonons has been linked to the thermal Hall conductivity~\cite{Grissonnanche2019}, ultrafast demagnetization~\cite{Choi2024, Tauchert2022}, and transient magnetic fields~\cite{Nova2017, Juraschek2022, Luo2023,Biggs2025_arXiv} harnessed for magnetic switching~\cite{Davies2024} or dynamical multiferroicity~\cite{Juraschek2017, Basini2024}. Nevertheless, the corresponding phonon trajectories, which would provide a direct observation of phonon angular momentum, could not be assessed in these studies, and the magnitude of their contribution to the observed magnetic effects remains debated~\cite{Merlin2024, Shabala2024, Merlin2025}.

To prepare a net angular momentum by the circular polarization of a doubly degenerate phonon mode, its orthogonal components need to be coherently driven with a phase delay of $\pi$/2. For IR-active $\text{E}_\text{u}$ modes, this can be achieved by a circularly polarized laser pulse in the terahertz (THz) spectral domain, where the resonant (linear) coupling between the THz electric field and the phonon’s electric dipole directly imprints the helicity of light onto the phonon mode. Yet, the resulting phonon angular momentum has only been inferred from secondary effects, such as phonon magnetic fields~\cite{Basini2024, Davies2024, Biggs2025_arXiv} or THz field-induced second harmonic generation \cite{Luo2023}. For Raman-active modes, the lattice trajectory can be controlled and directly observed by all-optical experiments that utilize stimulated Raman scattering techniques \cite{Wefers1998}. In all these examples, angular momentum was transferred from the light field to the lattice mode, and then potentially further to spin states, which are delicate to infer from magneto-optic measurements~\cite{Merlin2024, Merlin2025}. However, tracing the consecutive flow of angular momentum from the initially excited phonon to other lattice modes has remained elusive and represents a huge gap in our understanding of the ultrafast demagnetization sequence and other spin-lattice coupled phenomena since the pioneering experiments by Einstein, de Haas \cite{Einstein1915} and Barnett \cite{Barnett1915}.

\subsection*{Experimental concept}
Here, we provide experimental evidence for angular momentum $l_{\text{ph}}$ transfer between lattice modes by a coherent three-phonon scattering process, as sketched in Figure~1a. Exemplarily, we employ ionic Raman scattering, which is permitted by the lowest-order anharmonic lattice potential that couples two quanta of IR-active phonons to a Raman-active phonon. The topological insulator $\text{Bi}_2\text{Se}_3$ provides an ideal material platform to coherently drive such a three-phonon scattering process of axial phonons: Due to its inversion symmetry, $\text{Bi}_2\text{Se}_3$ accommodates either purely Raman- or IR-active doubly degenerate phonons in the $a$-$b$ plane parallel to the surface. At room temperature, an IR-active $\text{E}_\text{u}$ mode at 2.0~THz dominates the THz absorption~\cite{LaForge2010} (Extended Data Fig.~1). A Raman-active $\text{E}_\text{g}$ mode at 4.0~THz is anharmonically coupled to this 2~THz mode \cite{Melnikov2018}, perfectly fulfilling the double-frequency resonance condition $2\Omega^{\text{IR}}=\Omega^{\text{R}}$ (Methods) for sum-frequency ionic Raman scattering \cite{Juraschek2018, Blank2023}.

As shown in the experimental concept, sketched in Fig.~1b, we first resonantly excite a circularly polarized $\text{E}_\text{u}$ mode. In contrast to recent pioneering works \cite{Luo2023, Basini2024, Davies2024, Nova2017} employing such axial IR-phonons, we are not looking at the magneto-optic effects to study the angular momentum transfer to the spin system, instead we follow the trajectory of the coupled $\text{E}_\text{g}$ mode via the non-magnetic THz-induced Kerr effect (TKE)~\cite{Hoffmann2009, Johnson2019, Frenzel2023}. As both conservation of energy $2\hbar\Omega^{\text{IR}}=\hbar\Omega^{\text{R}}$ and linear momentum $2\hbar k^{\text{IR}}=\hbar k^{\text{R}}\approx 0$ (all $\Gamma$-modes) are perfectly fulfilled in this nonlinear process, we can study the isolated dynamics of angular momentum transfer and its conservation. This coherent upconversion-type coupling thus serves as an exemplary test bed to observe the fundamental mechanism of phonon-to-phonon angular momentum transfer, which can be generalized to any symmetry-allowed higher-linear momentum, higher-order or incoherent multi-phonon scattering processes~\cite{Kittel2005, Khalsa2024}. With this experimental design, we witness a helicity switch from one mode to the other, corresponding to the conservation of crystal angular momentum by a phononic analogue to a rotational Umklapp process~\cite{Bloembergen1980, Chen2015}.

\subsection*{Preparation and observation of coherent phonon angular momentum states}
We resonantly drive the $\text{E}_\text{u}$ mode of a $\text{Bi}_2\text{Se}_3$ single crystal epilayer (15~nm thickness), grown by molecular beam epitaxy on $\text{Al}_2\text{O}_3$~(Methods), with circularly polarized single-cycle THz pulses with peak fields exceeding 1~MV/cm (see Fig.~1c, blue shade). The normal incidence parallel to the three-fold symmetric c-axis of $\text{Bi}_2\text{Se}_3$ defines a common axis of rotation for consistent definition of photon and phonon helicities (Methods). We consecutively sample the transient birefringence induced by the in-plane oscillating coherent phonons with a sub-cycle - and thus phase-resolved - precision of a 20~fs probe pulse~(Methods). This time-domain probing scheme of coherent phonons via TKE is selectively sensitive to Raman-active modes~\cite{Maehrlein2017_PRL, Johnson2019, Frenzel2023}. With linear probe pulse polarizations at $45^\circ$ and $0^\circ$ in a laboratory frame defined by the THz field components $(E_x,E_y)$ measured at the sample position (Methods), we observe a strong TKE signal without signatures of transient magnetization (Supplementary Fig.~S8), but dominated by a long-lived oscillatory signal as shown in Fig.~1c. Its Fourier transform in Fig.~1d, unveils a single dominant oscillation at 4.0~THz, exactly at the $\text{E}_\text{g}$ mode's resonance~\cite{Melnikov2018}. Strikingly, this Raman-active phonon lies clearly outside the THz excitation spectrum (blue area in Fig.~1d) and thus must be nonlinearly excited either by the absorption of two photons~\cite{Maehrlein2017_PRL} or two coherent phonons \cite{Melnikov2018, Juraschek2018, Blank2023}, as confirmed by the quadratic field dependence of the $\text{E}_\text{g}$ amplitude in Extended Data Fig.~2b-e. Due to its lack of Raman activity, the $\text{E}_\text{u}$ mode is not observed by TKE, but instead by linear THz absorption (see pink feature in Fig.~1d taken from Extended Data Fig.~1), in full agreement with the mutual exclusion principle in the centrosymmetric $\text{Bi}_2\text{Se}_3$ crystal.

By measuring the probe polarization dependence of the phonon amplitude in Extended Data Fig.~3c, we show that the two orthogonal projections $Q_{\alpha}^{\text{R}}$ and $Q_{\beta}^{\text{R}}$ of the $\text{E}_\text{g}$ mode's amplitude $\boldsymbol{Q}^{\text{R}}$ can be separately measured by a  45° and 0° linear probe polarization, in full agreement with the two $\text{E}_\text{g}$-symmetry Raman tensors \cite{Richter1977}
\begin{equation}
\boldsymbol{R}_{\alpha} =  \begin{pmatrix} a & 0 \\ 0 & -a \end{pmatrix}, \quad
\boldsymbol{R}_{\beta} =  \begin{pmatrix} 0 & -a \\ -a & 0 \end{pmatrix}
\end{equation}
in the $\text{D}_{3\text{d}}$ point group, respectively~(see Methods). The corresponding microscopic projections also correspond to orthogonal in-plane ion motions, associated with "real" (as apposed to "pseudo") microscopic angular momentum, when driven with a $\pi/2$ phase difference (see Supplementary Text S1). We can thus trace the two dimensional (2D) lattice trajectory of the $\text{E}_\text{g}$ mode in time\cite{Wefers1998}. By implementing polarization-resolved electro-optic sampling in $\alpha$-quartz~\cite{Frenzel2024}, we consistently measure the full spatio-temporal trajectory of the THz electric field vector in the same plane, tracing a right-handed helicity in Fig.~1e. In the TKE signals in Fig.~1c, very similar oscillations are found at $45^\circ$ and $0^\circ$ probe polarization, respectively corresponding to the two orthogonal $\text{E}_\text{g}$ components $Q_{\alpha}^{\text{R}}$ and $Q_{\beta}^{\text{R}}$ with identical amplitude spectra in Fig.~1d. Both, the time domain inset in Fig.~1c and the complex Fourier transforms in Fig.~1d, unambiguously unveil a phase difference of exactly -$\pi$/2 between the orthogonal $Q_{\alpha}^{\text{R}}$ and $Q_{\beta}^{\text{R}}$ components. Therefore, we here demonstrate the first THz-driven preparation and simultaneous measurement of a coherent phonon angular momentum state, strictly termed axial (colloquially coined chiral) phonon~\cite{Juraschek2025_review}. This becomes directly evident from the 2D lattice trajectory in Fig.~1f, which strikingly reveals an opposite helicity compared to the driving field in Fig.~1e.

For systematic investigation of the phonon angular momentum transfer, we turn to tailored helical THz driving fields. By transformation of the exact THz field trajectories in Fig.~2a from time into frequency domain, and from a linear $(E_x,E_y)$ basis into a circular $(E_{|R\rangle},E_{|L\rangle})$ basis in Fig.~2b, we follow the complex evolution of the single cycle's helicity state across its broad spectrum. With this information, we customized a y-cut quartz waveplate plate (thickness 700~$\mu$m) to provide a pure right-hand (RCP) or left-hand circularly polarized (LCP) THz helicity state at 2~THz (see Fig.~2b). The three distinct fields with net helicity $l_\text{s}=-1,0,1$ at 2~THz in Fig.~2a, are then used to prepare the IR-active $\text{E}_\text{u}$ phonon in a prescribed coherent angular momentum state.

By comparison of the resulting $\text{E}_\text{g}$ phonon trajectories in Fig.~2c, we find perfectly opposite phonon helicities compared to the respective THz driving fields in Fig.~2a. Transforming also the phonon trajectories into frequency domain and circular basis, the $\text{E}_\text{g}$ phonons’ Stokes parameter $S_3(\Omega)=(|Q^{\text{R}}_{|R\rangle}(\Omega)|^2-|Q^{\text{R}}_{|L\rangle}(\Omega)|^2)/|Q^{\text{R}}(\Omega)|^2$ unveils pure helicities of two opposite angular momentum states across the full phonon bandwidth. Accordingly, we find a helicity switch: For the excitation with a RCP (LCP) field at 2~THz, the Raman-active phonon at 4~THz possesses a perfect LCP (RCP) and thus opposite helicity (arrows in Fig.~2a,c). In the following, we will address two arising questions: What is the nonlinear excitation mechanism of the observed phonon angular momentum state? And how is crystal angular momentum conserved leading to the switched helicity?

\subsection*{Pathways for angular momentum transfer}
Generally, the purely Raman-active $\text{E}_\text{g}$ mode can be driven by nonlinear photonic or nonlinear phononic pathways \cite{Maehrlein2017_PRL, Juraschek2018, Johnson2019} sketched in Fig.~3a. In the photonic case, pairs of photons within the broad THz excitation spectrum excite the $\text{E}_\text{g}$ mode via their sum-frequency polarization~\cite{Maehrlein2017_PRL}. For the nonlinear phononic coupling, the initially driven IR-active $\text{E}_\text{u}$ phonon $\boldsymbol{Q}^{\text{IR}}$ couples via an $(Q^{\text{IR}})^2Q^{\text{R}}$ anharmonic term in the lattice potential $V(\boldsymbol{Q}^{\text{IR}},\boldsymbol{Q}^{\text{R}})$ to the Raman-active mode, known as sum-frequency ionic Raman scattering \cite{Juraschek2018}. In the photonic case, the driving force for the $\text{E}_\text{g}$ mode is given by $\boldsymbol{E} {\cdot\boldsymbol{R}}_{\alpha} \boldsymbol{E}$ and $\boldsymbol{E} {\cdot\boldsymbol{R}}_{\beta} \boldsymbol{E}$ and is thus proportional to $E_x^2 - E_y^2$ and $-2E_xE_y$ for the $Q_{\alpha}^{\text{R}}$ and $Q_{\beta}^{\text{R}}$ component, respectively. In the $\text{D}_{3\text{d}}$ point group, to which $\text{Bi}_2\text{Se}_3$ belongs, the lowest-order anharmonic lattice potential that couples the modes of interest is $V(\boldsymbol{Q}^{\text{IR}},\boldsymbol{Q}^{\text{R}}) = c[(Q^{\text{IR}}_y)^2-(Q^{\text{IR}}_x)^2]Q^{\text{R}}_{\alpha}+2c\,Q^{\text{IR}}_x Q^{\text{IR}}_y Q^{\text{R}}_{\beta} + O(Q^4)$, where \textit{c} is the anharmonic coupling constant determined by our DFT calculations~(see Methods and Supplementary Text S2,3). Its derivative with respect to the Raman-active phonon coordinate $-\partial V/\partial \boldsymbol{Q}^{\text{R}}$ yields the ionic driving forces $F_{\alpha}=c[(Q^{\text{IR}}_x)^2-(Q^{\text{IR}}_y)^2]$ and $F_{\beta}=-2c\,Q^{\text{IR}}_x Q^{\text{IR}}_y$, respectively. Therefore, both driving mechanisms obey the same symmetry dictated by the $\text{D}_{3\text{d}}$ point group. One way to distinguish these two pathways is by their respective driving force dynamics, leading to different phonon dynamics when solving the respective $\boldsymbol{Q}^{\text{R}}$ equations of motion, as shown for $Q_{\alpha}^{\text{R}}$ in Fig.~3b. We find that the rise of the experimental trace, matching the $\text{E}_\text{u}$ lifetime as determined from THz transmission in Extended Data Fig.~1, is only reproduced by the phononic mechanism, pinpointing to the excited IR-active $\text{E}_\text{u}$ phonon as the primary driving force of the Raman-active phonon.

To quantify the contribution of the nonlinear phonon-phonon angular momentum transfer, we calculate the microscopic angular momentum $L_z^{\text{R}}$ of the excited $\text{E}_\text{g}$ phonon per unit cell from the photonic versus the phononic driving pathway. We compute all involved light-matter coupling parameters, Raman tensors, and anharmonic coupling parameters up to the 4th order using density functional theory~(DFT, see Methods). These DFT calculations and complementary symmetry considerations (Supplementary Text S1) attribute "real" angular momentum to both modes, independent of the azimuthal sample orientation and resulting from the real space atomic motions in our laboratory frame defined by the vectorial THz field~\cite{Streib2021} (Supplementary Movie S3). Strikingly, our \textit{ab-initio} model confirms a 3-order-of-magnitude larger angular momentum transfer from the $\text{E}_\text{u}$ mode to the $\text{E}_\text{g}$ mode compared to the direct driving of the $\text{E}_\text{g}$ mode via the photonic excitation process (see Fig.~3c). Taking into account the experimentally determined phonon coherence times of 1.6~ps and 1.1~ps for the IR- and Raman-active mode, respectively, and the measured THz field trajectory from Fig.~1e leads to a remarkable agreement with the measured phonon angular momentum dynamics $\boldsymbol{L}^{\text{R}}(t)=\boldsymbol{Q}^{\text{R}}(t) \times \dot{\boldsymbol{Q}}^{\text{R}}(t)$ in Fig.~3c. We find that $\sim$3\% of the absolute $\text{E}_\text{u}$ angular momentum is up-converted to the $\text{E}_\text{g}$ mode at the $\Gamma$-point, which strongly depends on the anharmonic coupling coefficient between the $\text{E}_\text{g}$ and $\text{E}_\text{u}$ mode, their frequency matching conditions and the angular momentum loss to the acoustic modes taken into account by the phenomenological damping constants. Accordingly, the observed coherent process inverts the usual equilibration flow of angular momentum and provides the first direct observation of anharmonic phonon-phonon angular momentum transfer. The remaining  $\text{E}_\text{u}$ angular momentum likely dissipates into lower-energy and higher-momentum acoustic modes, as sketched in Fig.~3c inset, which is generally permitted by an equivalent four-phonon scattering term $(Q^{\text{IR}})^2Q(-k)Q(k)$ of the anharmonic lattice potential \cite{Khalsa2024}. In the following we will discuss the underlying angular momentum conservation rules of the observed three-phonon scattering in the classical field and quantum particle picture.

\subsection*{Conservation of crystal angular momentum}
In the classical picture, the discrete rotational invariance of the crystal is captured by the point group symmetry and imprinted in the anharmonic lattice potential $V(\boldsymbol{Q}^{\text{IR}},\boldsymbol{Q}^{\text{R}})$. The lowest order symmetry-allowed squared-linear $(Q^{\text{IR}})^2Q^{\text{R}}$ coupling term thus dictates opposite helicities of the $\text{E}_\text{u}$ mode and its nonlinear driving force, as modeled in Fig.~4a,b, respectively. This becomes more apparent through transformation to a circular basis $Q^{\text{IR}}_{|R\rangle/|L\rangle} = (Q_x^{\text{IR}}\pm iQ_y^{\text{IR}})/\sqrt{2}$, which turns the driving force into $F_{|R\rangle/|L\rangle} = (F_{\alpha} \pm iF_{\beta})/\sqrt{2}$ and thus
\begin{equation}
F_{|R\rangle/|L\rangle}=c\,(Q^{\text{IR}}_{|L\rangle/|R\rangle})^2.
\end{equation}
From this, it is clear that the nonlinear phonon-phonon coupling doubles the $\text{E}_\text{u}$ frequency and inverts the helicity to enforce energy and angular momentum conservation, respectively (Fig.~4c).

Turning to the quasi-particle picture, we evoke the definition of phonon crystal (or pseudo) angular momentum (PAM) $l_\text{ph}$ as a conserved quantity under the $\text{C}_3$ rotational symmetry of $\text{Bi}_2\text{Se}_3$~\cite{Zhang2015}
\begin{equation} \label{PAMdef}
\hat{C}_3\phi_l(\varphi)=e^{i\frac{2\pi}{3}l_\text{ph}}\phi_l(\varphi),
\end{equation}
where $\phi_l$ is the wave function of the phonon at the high-symmetry point $k=0$. It is sufficient to depict the two orthogonal components of the  $\text{E}$ modes under $\text{C}_3$ symmetry on the simplest triangle lattice~\cite{Wefers1998}, as shown in Fig.~4d. As the axial mode is a superposition of two linearly polarized $\Gamma$-point modes, the $Q_{\alpha}$ and $Q_{\beta}$ components have identical phase relations at each of the three lattice sites. After performing an in-plane $\text{C}_3$ rotation (see Fig.~4e), an additional $2\pi/3$ phase factor has to be applied to obtain the initial eigenstate of Fig.~4d. In other words, the ions have to go a third of their oscillation period backwards in time to reach the same position as a $\text{C}_3$ rotation of a time snapshot. By definition in Eq.(\ref{PAMdef}), the PAM is thus $l_\text{ph}=-1$ per phonon for the observed LCP E mode. According to the same equation, in this $\text{C}_3$ symmetry $l_\text{ph}=-1$ is equivalent to $l_\text{ph}=2$, corresponding to an Umklapp process in a rotational Brillouin zone \cite{Bloembergen1980, Zhang2015,Chen2015}. Or simply speaking, as depicted in Fig.~4f, we may either move $4\pi/3$ of the phonon phase forward or $2\pi/3$ backwards in time to reconstruct the phonon wave function after the $\text{C}_3$ operation. In nonlinear phononics, the $\text{E}_\text{g}$ amplitude scales with the square of the $\text{E}_\text{u}$ driving mode, and thus inherits twice its phase: $Q^{\text{R}}\propto (Q^{\text{IR}})^2=e^{i2\phi_{\text{IR}}}(\hat{Q}^{\text{IR}})^2$. In the analogue particle picture, energy and PAM conservation of two annihilated $\text{E}_\text{u}$ phonon quanta with $l^{\text{IR}}_\text{ph}=1$ dictate that the $\text{E}_\text{g}$ phonon must carry twice the energy and PAM of the $\text{E}_\text{u}$ phonon, and thus $l^{\text{R}}_\text{ph}=2l^{\text{IR}}_\text{ph}=-1$. The experimentally observed exact reversal of phonon helicity would violate angular momentum conservation in free space (continuous rotational invariance), but perfectly obeys the crystal (or pseudo) angular momentum conservation under discrete $\text{C}_3$ rotational invariance in the investigated $ab$ plane of $\text{Bi}_2\text{Se}_3$. This phenomenon can be considered as rotational phonon-phonon Umklapp scattering, being the nonlinear \textit{phononics} analogue of the rotational Umklapp process in nonlinear optics~\cite{Bloembergen1980} and the \textit{angular} momentum counterpart of linear momentum phonon-phonon Umklapp scattering governing solid state physics~\cite{Kittel2005}.

\subsection*{General implications}
Most generally, our experimental observation confirms the long-standing hypothesis: Phonon-phonon angular momentum transfer is permitted via lattice anharmonicity and conserves crystal angular momentum. This has far reaching implications for spin-relaxation phenomena, such as ultrafast demagnetization, and eventually identifies one of the unknown intermediate steps of the (ultrafast) Einstein-de Haas effect~\cite{Einstein1915, Dornes2019}. Our findings suggest that "natural" incoherent angular momentum dissipation from higher energy to lower energy (and potentially higher linear momentum) modes follows anharmonic coupling mechanisms~\cite{Teitelbaum2018} equivalent to the anharmonic upconversion demonstrated in this work. This motivates higher momentum probing techniques like ultrafast electron~\cite{Waldecker2017, Stern2018} or X-ray~\cite{Teitelbaum2018} diffraction to study the incoherent relaxation of phonon-phonon angular momentum transfer.

The demonstrated coherent angular momentum and energy upconversion, moreover, opens the field of axial nonlinear phononics for selective control of chiral states of matter. Our observations in the topological insulator $\text{Bi}_2\text{Se}_3$ directly enable studies on phonon angular momentum-driven control of topological surface states via spin-selective electron-phonon scattering \cite{Braun2016a} and will stimulate novel investigations of phonon-driven topology switching \cite{Sie2019}. Furthermore, the witnessed rotational phonon-phonon Umklapp scattering may play a similarly fundamental role for angular momentum dissipation and thermal Hall conductivity \cite{Grissonnanche2019}, as the linear momentum Umklapp processes does  in heat transport \cite{Kittel2005}. Our observation of back-folded PAM states motivates further studies on phonon rotational Bloch functions and phonon driven Floquet matter~\cite{Huebener2018, Schultheiss2025_arXiv}, with potential implications on topological phonon states \cite{Xu2024}, active control of spin-valley polarization \cite{Zhu2018}, or phonon orbital angular momentum states \cite{Gao2023}.

In conclusion, our results experimentally confirm the full analogy to energy and linear momentum relaxation in solids: The anharmonicity of the crystal lattice mediates the transfer and conservation of phonon angular momentum. In the classical picture, the discrete rotational symmetry imprinted in the anharmonic lattice potential enforces the observed helicity switch between the two coupled modes. In the quasi-particle picture, crystal angular momentum is transferred and conserved by a rotational Umklapp process during three-phonon scattering. This concept can be extended to any higher-order lattice anharmonicity and multi-phonon scattering abundant in nature, and therefore completes the solid-state physics picture of axial and chiral phonons as quasi-particles responsible for carrying and conserving crystal angular momentum. In future, the here established axial nonlinear phononics will provide a precise handle for ultrafast control over spins, topologies and chiral quasi-particles.

\clearpage 

\subsection*{Acknowledgments}
We thank Michael Fechner, Hiroki Ueda, Prakriti P. Joshi, Alexander Paarmann, Ralph Ernstorfer and Alfred Leitenstorfer for fruitful and friendly discussions. S.F.M. acknowledges support and funding from the Deutsche Forschungsgemeinschaft (DFG, grant No. 469405347) for his Emmy Noether group “Circular Phononics”. D.M.J. acknowledges support from the Israel Science Foundation (ISF) Grant No. 1077/23 and 1916/23. D.M.J. further acknowledges support from the ERC Starting Grant CHIRALPHONONICS, no. 101166037.
\subsection*{Author contributions}
S.F.M. conceived the experimental design and supervised the project. M.F., M.S.S., and J.M.U. additionally contributed experimental ideas. O.M. and M.F. performed the main experiments. M.S.S. performed the THz transmission measurements. C.R. and G.M. synthesized and characterized the samples. O.M., M.F., M.S.S., J.M.U., and S.F.M. analyzed and interpreted the experimental data. O.M., M.S.S., and J.M.U. conceived and performed the classical equations of motion. C.P. and D.M.J. performed the \textit{ab-initio} and semi-classical model calculations. O.M. and M.S.S. carried out the symmetry analysis. The manuscript was written by S.F.M., O.M., C.P., and D.M.J., with contributions from all co-authors.
\subsection*{Competing interests}
The authors declare that they have no competing interests.

\clearpage
\begin{figure} 
	\centering
	\includegraphics[width=0.65\textwidth]{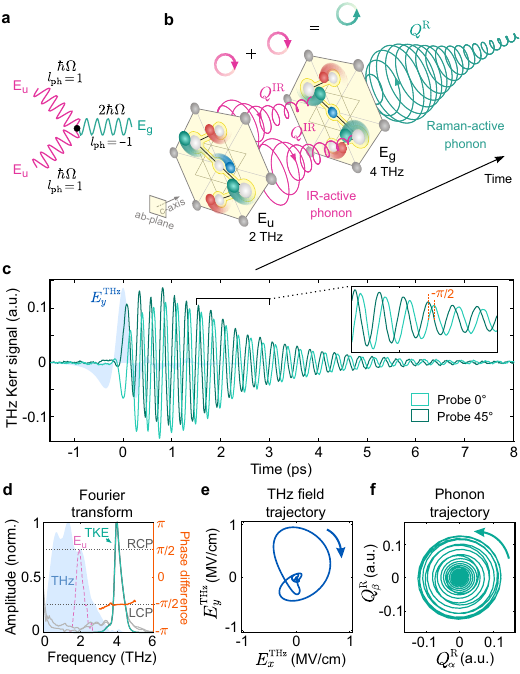} 
	\caption{\textbf{Phonon-to-phonon angular momentum transfer in bismuth selenide.}
    \textbf{a}, Investigated three-phonon scattering process: annihilation of two $\text{E}_\text{u}$ phonons producing one $\text{E}_\text{g}$ phonon must conserve energy, (pseudo) linear momentum and angular momentum. \textbf{b}, Experimental concept: A coherent circular $\text{E}_\text{u}$ mode (pink) at 2~THz nonlinearly transfers energy and momenta to the circular $\text{E}_\text{g}$ mode (green) at 4~THz with opposite helicity in $\text{Bi}_2\text{Se}_3$. \textbf{c}, THz-induced Kerr effect (TKE) in $\text{Bi}_2\text{Se}_3$ (green) under right-handed elliptical THz excitation (blue projection) for two linear probe pulse polarizations ($0^\circ$ and $45^\circ$) measuring two orthogonal $\text{E}_\text{g}$ phonon components. Inset: zoom into time domain traces. \textbf{d}, Fourier transforms unveil equal amplitudes (green), but perfect $-\pi/2$ phase difference (orange) of the $\text{E}_\text{g}$ phonon projections in \textbf{c}, clearly outside the excitation spectrum (blue shade) and at twice the $\text{E}_\text{u}$ phonon resonance (pink). \textbf{e}, Trajectory of the THz excitation pulse’s electric field vector. \textbf{f}, $\text{E}_\text{g}$ phonon trajectory obtained by Fourier-filtered TKE signal from \textbf{c} unveils opposite helicity with respect to the THz field in \textbf{a}.}
	\label{fig:main_1} 
\end{figure}

\begin{figure} 
	\centering
	\includegraphics[width=\textwidth]{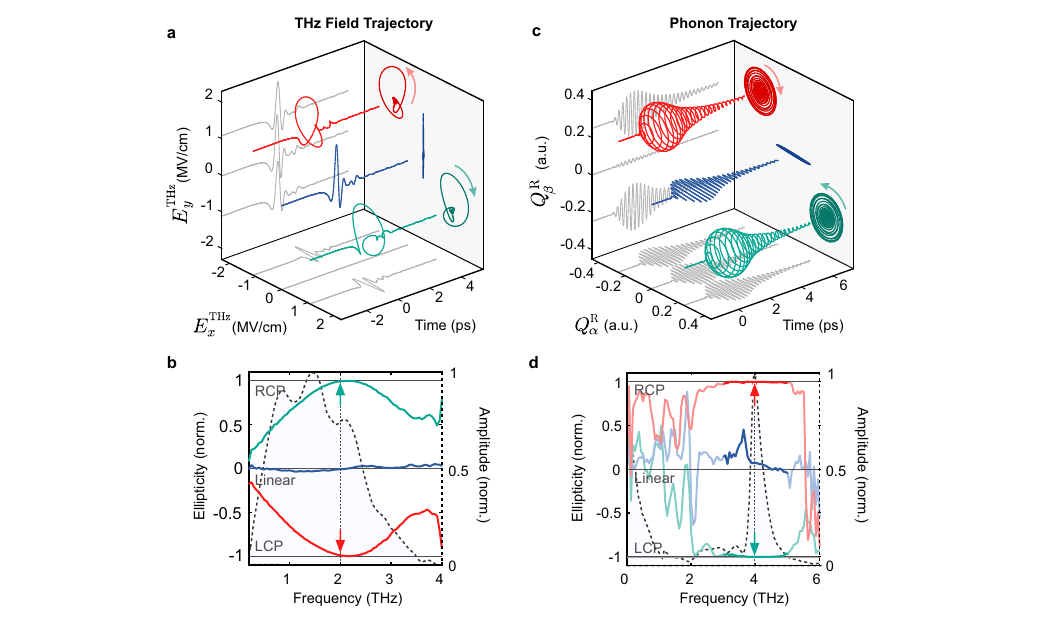} 
	\caption{\textbf{THz field and phonon trajectories and their helicity states.}
    \textbf{a}, Polarization-resolved EOS of the THz excitation electric fields tailored by a 700~$\mu$m-thick y-cut quartz plate at three different azimuthal orientations. \textbf{b}, Frequency-decomposed ellipticity (colored lines) and amplitude (dashed black line) of the excitation fields in \textbf{a}, calculated from the Stokes parameter $S_3$ of the THz field’s Fourier transform. At around 2~THz, the THz pump pulses exhibit well-defined linear (blue), right-circular (green), and left-circular (red) polarizations. \textbf{c}, Measured $\text{E}_{\text{g}}$ phonon trajectories for each THz polarization state from \textbf{a}. \textbf{d}, Corresponding ellipticity decompositions of the TKE signals (colored lines) and Fourier filtered $\text{E}_{\text{g}}$ trajectories in \textbf{c} (darker colors). The $\text{E}_{\text{g}}$ phonon at 4~THz shows full helicity reversal with respect to the THz electric field at 2~THz. The dashed black line represents the TKE spectral amplitude.}
	\label{fig:main_2} 
\end{figure}

\begin{figure} 
	\centering
	\includegraphics[width=0.61\textwidth]{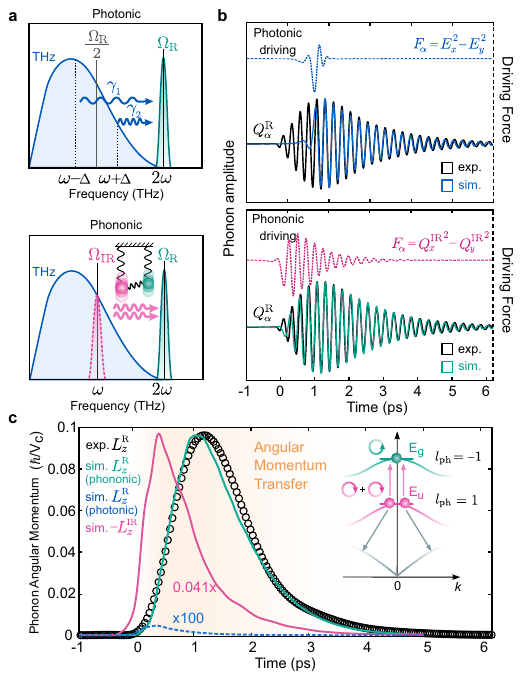} 
	\caption{\textbf{Angular momentum transfer channels and dynamics.}
    \textbf{a}, Schematic nonlinear excitation pathways: "Photonic" THz sum-frequency excitation (top) involves two spectral THz components driving a Raman-active phonon at their sum frequency. "Phononic" sum-frequency ionic Raman scattering (bottom) involves a resonantly driven IR-active phonon ($\text{E}_{\text{u}}$) coupling nonlinearly to a Raman-active mode ($\text{E}_{\text{g}}$). \textbf{b}, Modeled driving forces $F_{\alpha}$ (dotted lines) and resulting phonon amplitudes $Q_{\alpha}^{\textrm{R}}$ (blue, green) for photonic (top; offset in time to match phonon decay dynamics) and phononic (bottom) pathways compared to experimental data (black) under right-handed circular THz excitation. \textbf{c}, Quantitative \textit{ab-initio} modeling of angular momentum dynamics for all involved modes and pathways. Experimentally measured angular momentum dynamics $L_z^{\textrm{R}}$ (black circles) are normalized to the peak magnitude of the calculated phononic pathway $L_z^{\textrm{R}}$ (green). The shaded orange region highlights angular momentum transfer from $\text{E}_{\text{u}}$ (pink) to $\text{E}_{\text{g}}$, resulting in a delayed rise time and three orders of magnitude higher angular momentum transfer (green) compared to photonic pathway (blue). Inset: ~3\% of the $\text{E}_{\text{u}}$ angular momentum is transferred to the $\text{E}_{\text{g}}$ mode, the rest likely dissipates into lower-energy and higher-momentum modes (gray).}
	\label{fig:main_3} 
\end{figure}

\begin{figure} 
	\centering
	\includegraphics[width=0.9\textwidth]{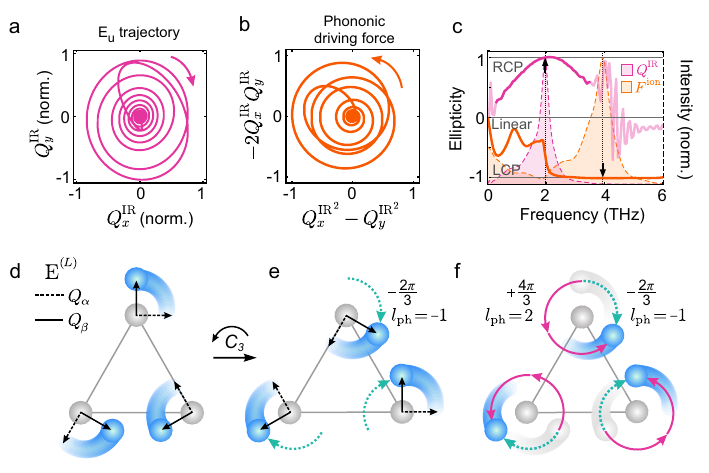} 
	\caption{\textbf{Conservation of phonon angular momentum.}
    \textbf{a}-\textbf{c}, Classical field picture: Helicity reversal dictated by the symmetry of anharmonic lattice potential. \textbf{a}, $\text{E}_{\text{u}}$ phonon trajectory calculated from the experimental RCP THz excitation field. \textbf{b}, Corresponding phononic driving force for the $\text{E}_{\text{g}}$ mode, rotating in the opposite direction. \textbf{c}, Frequency-decomposed ellipticity of the $\text{E}_{\text{u}}$ phonon $Q^{\textrm{IR}}$ (pink) and its phononic driving force $F^{\textrm{ion}}$ (orange) showing opposite helicities at 2 and 4~THz, respectively. \textbf{d}-\textbf{f}, PAM conservation in the particle picture under three-fold rotational invariance. \textbf{d}, Solid and dashed black arrows indicate the orthogonal components of a doubly-degenerate E-mode in a triangular lattice. All lattice sites are in phase in the case of a $\Gamma$-mode. \textbf{e}, Under three-fold rotation, the crystal remains invariant up to a phase factor of $2\pi l_{\textrm{ph}}/3$, which defines the PAM to $l_{\textrm{ph}}=-1$ here. \textbf{f}, Equivalence of $l_{\textrm{ph}}=-1$ and $l_{\textrm{ph}}=+2$ in three-fold symmetric groups. $l_\textrm{ph}=+2$ is provided by the two-phonon absorption, corresponding to a rotational three-phonon Umklapp scattering.}
	\label{fig:main_4} 
\end{figure}


\clearpage 

%

%
%
%
%
%
%


\section*{Acknowledgments}
We thank Michael Fechner, Hiroki Ueda, Prakriti P. Joshi, Leona Ocean Nest, Alexander Paarmann, Tobias Kampfrath, Ralph Ernstorfer and Alfred Leitenstorfer for fruitful and friendly discussions. S.F.M. acknowledges support and funding from the Deutsche Forschungsgemeinschaft (DFG, grant No. 469405347) for his Emmy Noether group “Circular Phononics”. D.M.J. acknowledges support from the Israel Science Foundation (ISF) Grant No. 1077/23 and 1916/23. D.M.J. further acknowledges support from the ERC Starting Grant CHIRALPHONONICS, no. 101166037.

\section*{Author contributions}
S.F.M. conceived the experimental design and supervised the project. M.F., M.S.S., and J.M.U. additionally contributed experimental ideas. O.M. and M.F. performed the main experiments. M.S.S. performed the THz transmission measurements. C.R. and G.M. synthesized and characterized the samples. O.M., M.F., M.S.S., J.M.U., and S.F.M. analyzed and interpreted the experimental data. O.M., M.S.S., and J.M.U. conceived and performed the classical equations of motion. C.P. and D.M.J. performed the \textit{ab-initio} and semi-classical model calculations. O.M. and M.S.S. carried out the symmetry analysis. The manuscript was written by S.F.M., O.M., C.P., and D.M.J., with contributions from all co-authors.

\section*{Competing interests}
The authors declare no competing interests.

\section*{Methods}
\subsection*{Sample details}\label{Msubsec1}
The sample was grown on a sapphire (0001) substrate by means of molecular-beam epitaxy. Prior to the growth of the $\text{Bi}_2\text{Se}_3$ epilayer, the sapphire substrate was annealed at 900°C for 30 minutes. Subsequently, the substrate temperature was reduced to 270°C for the growth of the $\text{Bi}_2\text{Se}_3$ film, which took place in a Se overpressure regime. The Bi and Se partial pressures amount to $5.3\times 10^{-8}$~mbar and $5.8\times 10^{-7}$~mbar, respectively, which results in a growth rate of 18.3~nm/h. 

The sample was ex-situ characterized by x-ray diffraction, employing a Rigaku SmartLab XRD system. The film thickness was determined by means of x-ray reflectivity, shown in Fig.~S5. Numerous oscillations are seen in the XRR curve, evidencing the high surface and interface quality of the $\text{Bi}_2\text{Se}_3$ film. From the periodicity of the oscillations, a $\text{Bi}_2\text{Se}_3$ film thickness of 15.0~nm is determined. To obtain information about the epilayer's crystal structure, XRD measurements were performed (Fig.~S6). Besides the three peaks stemming from the sapphire substrate, a total of seven (0,0,l) peaks from the $\text{Bi}_2\text{Se}_3$ epilayer are seen, proving the single-crystal nature of the $\text{Bi}_2\text{Se}_3$ film with the c-axis in growth direction.

\subsection*{Experimental setup}\label{Msubsec2}
The experimental setup is shown in Extended Data Fig.~2a. Intense single-cycle THz pulses (1.5~THz central frequency, 2.0~THz spectral FWHM, 1~kHz repetition rate) with peak fields exceeding 1.5~MV/cm are generated by optical rectification in $\text{LiNbO}_3$ using the tilted pulse front technique \cite{Hirori2011}. Synchronized optical probe pulses (790~nm central wavelength, 1.6~nJ pulse energy, 80~MHz repetition rate, 20~fs pulse duration) are collinearly incident on the sample at normal incidence. The probe's linear polarization is controlled using a half-wave plate to resolve two orthogonal components of the $\text{E}_\text{g}$ phonon mode (Fig.~1f). The intense THz pulses induce a transient birefringence in the sample, which alters the polarization of the transmitted optical probe pulses (THz-induced Kerr effect, TKE \cite{Hoffmann2009}). Coherently excited Raman-active phonons also contribute to this transient birefringence by modulating the linear susceptibility \cite{Frenzel2023,  Urban2025}. A balanced detection scheme comprising a half-wave plate, Wollaston prism, and balanced photodiodes is used to measure this change in polarization.

\subsection*{Helicity-tailored THz fields}\label{Msubsec3}
To control the THz polarization state, we use a birefringent 700~$\mu$m-thick y-cut quartz waveplate that imparts a frequency-dependent relative phase retardation between the orthogonal fast and slow crystalline axes (Fig.~S7a,b). Due to the highly polychromatic nature of a single-cycle THz pulse, the waveplate generates a continuum of elliptical polarization states across the broad spectrum constituting the pulse, but is tailored to have a particular thickness to generate a desired circular polarization state at 2~THz, where the $\text{E}_\text{u}$ phonon resonance is situated. By orienting the quartz waveplate to 0°, -45°, +45° with respect to the vertical incident THz polarization, the transmitted THz polarization state is toggled between linear, right-, and left-circular polarizations at 2~THz (Extended Data Fig.~4a,b). In addition, to investigate arbitrary lattice trajectories (Extended Data Fig.~5l), we use a 380~$\mu$m-thick y-cut quartz waveplate (Fig.~S7c,d) generating elliptically polarized THz beam (Extended Data Fig.~5a,b).

\subsection*{Measuring $\text{E}_\text{u}$ phonon lifetime via THz transmission}\label{Msubsec4}
We perform THz time-domain spectroscopy in a double-focus geometry, shown in Extended Data Fig.~1a, to measure the THz transmittance of a $\text{Bi}_2\text{Se}_3$ film. A spintronic THz emitter (TeraSpinTech T-Spin2) generates an ultrabroadband THz pulse~\cite{Rouzegar2023}, which is focused onto the sample in the first focal position. The transmitted THz pulse is then refocused onto a 30~$\mu$m-thick $\text{GaP}$ crystal for electro-optic sampling. To determine the intrinsic THz transmittance $T_{\text{Bi}_2\text{Se}_3}(\omega)$ of $\text{Bi}_2\text{Se}_3$, we conduct two measurements: (1) the THz transmission of a 0.3~$\mu$m-thick $\text{Bi}_2\text{Se}_3$ film grown on a 500~$\mu$m sapphire substrate and (2) a reference THz transmission through an identical 500~$\mu$m-thick sapphire wafer (Extended Data Fig.~1b,c). Taking the ratio of these spectra eliminates the detector response function of the $\text{GaP}$ crystal, and partially removes the substrate contribution, and therefore contains the intrinsic THz transmittance properties of $\text{Bi}_2\text{Se}_3$ (Extended Data Fig.~1d). The resultant characteristic absorption feature at 2~THz confirms the dominant absorption due to the resonant excitation of the IR-active $\text{E}_\text{u}$ phonon mode\cite{LaForge2010}. 

We identify an analytical expression for the THz transmission function $T_{\text{Bi}_2\text{Se}_3}(\omega)$ based on Fresnel coefficients, incorporating the $\text{Bi}_2\text{Se}_3$ dielectric function to extract the $\text{E}_\text{u}$ phonon frequency $\omega_{\text{IR}}$ and damping rate $\gamma_{\text{IR}}$, which defines the phonon lifetime $\tau_{\text{IR}}=1/(2\pi\gamma_{\text{IR}})$. 
The derived transmission function $T_{\text{Bi}_2\text{Se}_3}(\omega)$ (see Supplementary Text S4 for the derivation) accounts for multiple internal reflections within $\text{Bi}_2\text{Se}_3$, closely matching the measured $\text{Bi}_2\text{Se}_3$ transmittance (Extended Data Fig.~1d). By modeling the experimental data, we determine the $\text{E}_\text{u}$ phonon parameters to be $\omega_\text{IR}$=1.97~THz and $\gamma_\text{IR}$=0.2~THz, corresponding to the phonon lifetime $\tau_\text{IR}$=0.79~ps. 

\subsection*{Helicity state characterization}\label{Msubsec5}
In our experimental geometry, the co-propagating pump and probe pulses arrive at normal incidence on the $\text{Bi}_2\text{Se}_3$ surface, establishing a unique optical axis that coincides with the crystallographic $c$-axis of the hexagonal unit cell (confimed by XRD, Fig.~S6) and thus with its $C_3$ symmetry axis. The transverse THz electric field components, $E_x(t)$ and $E_y(t)$, measured with full sign sensitivity at the sample position~\cite{Frenzel2024}, define the laboratory reference frame and its handedness. This configuration ensures a single, well-defined rotation axis for all polarization analysis, allowing the helicity of both the THz field and the coherently driven $\text{E}_\text{g}$ phonon to be characterized and compared within the same coordinate system.

We characterize the complex helicity states of the single-cycle THz pulses (Fig.~2b) and arbitrarily polarized phonon states (Fig.~2d) by computing the ellipticity parameter in the frequency domain. The ellipticity, which quantifies the relative contribution of circular polarization components, is calculated as the normalized third Stokes parameter $S_3(\omega)$, independently evaluated at each frequency. To achieve this, the electric field components $E_x(\omega)$ and $E_y(\omega)$ in the Cartesian basis $(\hat{x},\hat{y})$ are transformed into the circular basis $(\hat{R},\hat{L})$. The circular basis is defined according to the standard optics convention for right- and left-circular polarization, and the sign of the phonon pseudo angular momentum in Eq.~(3) is chosen accordingly. The corresponding field components in the circular basis are computed as $E_R(\omega)=(E_x(\omega)+iE_y(\omega))/\sqrt{2}$ and $E_L(\omega)=(E_x(\omega)-iE_y(\omega))/\sqrt{2}$. The ellipticity parameter $S_3(\omega)$ is then determined as $S_3(\omega)=(|E_R(\omega)|^2-|E_L(\omega)|^2)/|E(\omega)|^{\,2}$, where $|E(\omega)|^{\,2} = |E_R(\omega)|^2+|E_L(\omega)|^2$. The parameter takes values in the range $[-1,1]$, with $S_3(\omega)$=1 corresponding to purely right-handed circular polarization (RCP), $S_3(\omega)=-1$ to purely left-handed circular polarization (LCP), and $S_3(\omega)=0$ to any possible linear polarization. The same procedure is applied to characterize the polarization state of a phonon with the amplitude $\boldsymbol{Q}^{\text{R}}(\omega)=(Q^{\text{R}}_{\alpha}(\omega),Q^{\text{R}}_{\beta}(\omega))$. This approach provides a frequency-resolved description of the helicity states for both THz pulses and phonon modes.

\subsection*{Probing $\text{E}_\text{g}$ phonon trajectories}\label{Msubsec6}

The coherent phonon dynamics of the $\text{E}_\text{g}$ mode are monitored via the THz-induced Kerr effect, where the intense THz pulses induce transient optical birefringence in $\text{Bi}_2\text{Se}_3$ sample, mediated by the coherent lattice vibration. As the optical probe pulse propagates through the sample, its polarization undergoes a rotation proportional to the instantaneous amplitude of the coherent phonon. We note that the temporal evolution of the $\text{E}_\text{g}$ mode is thus measured by varying the time delay between the THz pump and optical probe pulses. 

Importantly, the THz-induced Kerr effect originating from coherent phonons involves a Raman-type probing mechanism~\cite{Urban2025}, in which the THz pump and optical probe fields couple via the third-order nonlinear susceptibility, analogous to stimulated Raman scattering experiments. In this regard, the Raman tensor formalism provides a framework for understanding the probing of phonon trajectories.

The probing signal $S^\text{pr}$ of a particular phonon's amplitude is dependent on the in-plane azimuthal angle of the probe pulse polarization, and the signal amplitude is determined by the phonon symmetry encoded in the Raman tensor. This Raman-type probing via transient birefringence can be expressed as $S^\text{pr}(\theta)\propto \hat{e}_i^{\perp\text{pr}}(\theta)Q(t)R^Q_{ij}\hat{e}_j^{\text{pr}}(\theta)$, where $\hat{e}^{\text{pr}}(\theta)$ and $\hat{e}^{\perp\text{pr}}(\theta)$ are the polarization vector of the incident probe pulse oriented at the angle $\theta$ in the laboratory frame, and a unit vector perpendicular to it, respectively. $R^Q_{ij}$ is the Raman tensor of a given phonon mode $Q(t)$.

In the case of a doubly degenerate mode, such as the $\text{E}_\text{g}$ mode, two distinct Raman tensors are involved in the probing process, corresponding to the two orthogonal components that form a basis for polarization-resolved detection. In this representation, the probing signal is given by $S^\text{pr}\propto \hat{e}_i^{\perp\text{pr}}(Q_{\alpha}(t)R^{Q_{\alpha}}_{ij}+Q_{\beta}(t)R^{Q_{\beta}}_{ij})\hat{e}_j^{\text{pr}}(\theta)$, where $Q_{\alpha}(t)$ and $Q_{\beta}(t)$ are the classical amplitudes of the orthogonal phonon components, $R^{Q_{\alpha}}_{ij}$ and $R^{Q_{\beta}}_{ij}$ are the corresponding Raman tensors in the laboratory frame $(\hat{x},\hat{y})$. For the $\text{E}_\text{g}$ phonon mode, the Raman tensors of two orthogonal projections $Q_{\alpha}(t)$ and $Q_{\beta}(t)$ are given by \cite{Richter1977, Loudon1964}
\begin{equation}
\boldsymbol{R}_{\alpha} =  \begin{pmatrix} a & 0 \\ 0 & -a \end{pmatrix}, \quad
\boldsymbol{R}_{\beta} =  \begin{pmatrix} 0 & -a \\ -a & 0 \end{pmatrix}.
\label{Raman_tensors}
\end{equation}
Therefore, for a probe pulse with an incident polarization $\hat{e}^{\text{pr}}=(-\sin{\theta}, \cos{\theta})$, which is aligned vertically at $\theta$= 0°, the resulting probing signal is given by $S^\text{pr}(t, \theta) \propto Q_{\alpha}(t)\sin{2\theta}+Q_{\beta}(t)\cos{2\theta}$. This angular dependence reflects the quadrupolar symmetry of the $\text{E}_\text{g}$ phonon mode, and enables selective detection of its orthogonal components. By aligning the probe polarization to $\theta$=45° and $\theta$=0°, the contribution of $Q_{\alpha}(t)$ and $Q_{\beta}(t)$ can be isolated (Extended Data Fig.~3c), respectively, allowing for the reconstruction of the $\text{E}_\text{g}$ phonon trajectory~\cite{Wefers1998}.

Note that all TKE and polarization-resolved EOS measurements use the same probe polarizations to resolve the x- and y-components of the $\text{E}_\text{g}$ phonon state and THz pump field. This results from the shared three-fold rotational symmetry property of $\text{Bi}_2\text{Se}_3$ ($\text{D}_{3\text{d}}$ point group) and the $\alpha$-quartz EOS detector crystal ($\text{D}_{3}$ point group). The EOS signal $S^\text{EOS}(t, \theta) \propto E^{\text{THz}}_x(t)\sin{2\theta}+E^{\text{THz}}_y(t)\cos{2\theta}$ follows directly from the second-order nonlinear polarization $\boldsymbol{P}^{(2)}$ defined by the $\chi^{(2)}$ tensor of $\text{D}_{3}$ symmetry~\cite{Frenzel2024}. $S^\text{EOS}$ and $S^\text{pr}$ exhibit the same angular dependence, ensuring that the polarizations of both the $\text{E}_\text{g}$ phonon and the THz field are measured consistently in the same reference frame, enabling unambiguous identification of the helicity reversal.

\subsection*{Excitation pathways of the $\text{E}_\text{g}$ mode }\label{Msubsec7}
The higher-frequency, Raman-active $\text{E}_\text{g}$ phonon mode can be driven via two distinct pathways: photonic sum-frequency excitation~\cite{Maehrlein2017_PRL} and sum-frequency ionic Raman scattering~\cite{Juraschek2018} through anharmonic coupling with IR-active $\text{E}_\text{u}$ phonon. The temporal evolution of the doubly-degenerate $\text{E}_\text{g}$ phonon amplitude can be described classically by a system of two equations of motion with the vectorial driving force $\boldsymbol{F}(t)$
\begin{equation} \label{eqmotion}
 \ddot{Q}^{\text{R}}_{i} + \gamma_{\text{R}} \dot{Q}^{\text{R}}_{i} + \omega^2_{\text{R}} Q^{\text{R}}_i = F_i(t), \quad i=\{{\alpha},{\beta}\},
\end{equation}
where $\boldsymbol{Q}^{\text{R}}$ is the coherent amplitude of the $\text{E}_\text{g}$ mode. 

In the photonic pathway, the THz field $\boldsymbol{E}^{\text{THz}}=(E_{\text{x}}, E_{\text{y}})$ directly couples to the Raman tensors of the $\text{E}_\text{g}$ mode, yielding a driving force
\begin{equation}\label{Force_photonic}
\boldsymbol{F}^{\text{ph}}_{\text{R}} \propto \begin{pmatrix} R^{ij}_{\alpha}E_iE_j \\ R^{ij}_{\beta}E_iE_j \end{pmatrix} = a
\begin{pmatrix}
E_x^2 - E_y^2 \\ -2E_xE_y
\end{pmatrix}.
\end{equation}

In the phononic pathway, the driving force originates from the anharmonic potential of the Raman-active $\text{E}_\text{g}$ mode and the IR-active $\text{E}_\text{u}$ mode at half its frequency. The lowest-order, symmetry-allowed anharmonic potential coupling the Raman-active $\text{E}_\text{g}$ and the IR-active $\text{E}_\text{u}$ modes is given by
\begin{equation}
V(\boldsymbol{Q}^{\text{IR}},\boldsymbol{Q}^{\text{R}}) = c[(Q^{\text{IR}}_y)^2-(Q^{\text{IR}}_x)^2]Q^{\text{R}}_{\alpha}+2cQ^{\text{IR}}_x Q^{\text{IR}}_y Q^{\text{R}}_{\beta} + O(Q^4),
\label{eq:anharmonic_potential}
\end{equation}
where $\boldsymbol{Q}^{\text{IR}}$ and $\boldsymbol{Q}^{\text{R}}$ are coherent phonon amplitudes of $\text{E}_\text{u}$ and $\text{E}_\text{g}$ phonon states, respectively. This simplest form of the potential results in a phononic driving force on the $\text{E}_\text{g}$ mode
\begin{equation}
\boldsymbol{F}^{\text{ion}}_{\text{R}} \propto \begin{pmatrix} -\partial V/\partial Q_{\alpha}^{\text{R}} \\ -\partial V/\partial Q_{\beta}^{\text{R}}  \end{pmatrix} = c
\begin{pmatrix}\label{Force_phononic}
(Q_x^{\text{IR}})^2 - (Q_y^{\text{IR}})^2 \\ -2Q_x^{\text{IR}}Q_y^{\text{IR}}
\end{pmatrix}.
\end{equation}

Spectrally, the phononic driving force $\boldsymbol{F}^{\text{ion}}_{\text{R}}(\omega)$ is given by the convolution of two Lorentzians $Q^{\text{IR}}(\omega)$ centered each at $\Omega_{\text{IR}}$, yielding a spectrum centered at $2\Omega_{\text{IR}}$ with linewidth $2\gamma_{\text{IR}}$ (see Fig.~4c) The Raman-active mode is efficiently driven via the anharmonic $Q^2_{\text{IR}} Q_{\text{R}}$ coupling when its frequency $\Omega_{\text{R}}$ overlaps with the spectral window of the driving force, reaching maximum coupling at $\Omega_{\text{R}} = 2\Omega_{\text{IR}}$. If this frequency window is not matched, higher-order anharmonic terms may still permit the coupling.

The photonic (Eq.~\ref{Force_photonic}) and phononic (Eq.~\ref{Force_phononic}) driving forces exhibit identical spatio-rotational symmetry, as the THz field and the IR-active $\text{E}_\text{u}$ mode both transform as a vector representation within the $\text{D}_{3\text{d}}$ point group of the $\text{Bi}_2\text{Se}_3$ crystal lattice. While the forces’ spatial symmetries are the same, their temporal profiles differ: the photonic driving force follows the duration of the single-cycle THz pulse, whereas the phononic driving force is governed, for short THz excitation pulses, by the lifetime of the IR-active $\text{E}_\text{u}$ phonon, setting the timescale for phonon-phonon coupling (Fig.~3b).

The different durations of the driving forces result in distinctive rise times in the $\text{E}_\text{g}$ phonon evolution. To identify the dominant excitation mechanism, we analyze the rise time of the measured $\text{E}_\text{g}$ phonon dynamics and model both excitation pathways. For the photonic driving case, we calculate the $\text{E}_\text{g}$ phonon evolution $\boldsymbol{Q}^{\text{R}}(t)$ by solving the equations of motion (Eq.~\ref{eqmotion}) with the driving force defined in Eq.~\ref{Force_photonic}, using the electric field components of the circularly polarized THz field measured by polarization-resolved EOS (Fig.~1e). We focus on the temporal dynamics, rather than the absolute amplitudes, and therefore set the Raman tensor coefficient $a$ to unity. The damping rate $\gamma_{\text{R}}$ used here is identified from the TKE signal and equals 0.22~THz.

To model the phononic case, we solve the system of four equations corresponding to the polarization components of the $\text{E}_\text{u}$ and $\text{E}_\text{g}$ phonon modes $Q=\{Q_x^{\text{IR}}, Q_y^{\text{IR}}, Q_{\alpha}^{\text{R}}, Q_{\beta}^{\text{R}}\}$. Here, the IR-active mode is linearly driven by the same circularly-polarized THz pulse (Fig.~1e) and coupled with the Raman-active mode via the anharmonic potential $V(\boldsymbol{Q}^{\text{IR}},\boldsymbol{Q}^{\text{R}})$, so its driving force's components are defined as $F^{\text{IR}}_i = ZE_i^{\text{THz}} - \partial V/\partial Q_i^{\text{IR}}$. The Born effective charge $Z$ and phonon coupling parameter $c$ are set to unity for simplicity. The driving force acting on the $\text{E}_\text{g}$ mode is purely defined by the anharmonic coupling potential $F^{\text{R}}_i = - \partial V/\partial Q_i^{\text{R}}$. The parameters for the $\text{E}_\text{u}$ phonon mode ($\omega_{\text{IR}}$=1.97~THz, $\gamma_{\text{IR}}$=0.20~THz) are obtained from THz transmission measurements (see Methods and Supplementary Text S4) and the $\text{E}_\text{g}$ phonon damping rate is identified by matching the TKE signal decay dynamics ($\gamma_{\text{R}}$=0.29~THz, corresponding to the phonon lifetime $\tau_{\text{R}}=1/(2\pi\gamma_{\text{R}}$)= 0.55~ps).

\subsection*{First-principles calculations}\label{Msubsec8}
We calculate the phonon eigenfrequencies, eigenvectors, Born effective charges and dielectric function of $\text{Bi}_2\text{Se}_3$ using the density functional theory formalism as implemented in \textsc{vasp}, as well as the frozen-phonon method as implemented in \textsc{phonopy}. We use PAW pseudopotentials with valence electron configurations Bi($6s^2 6p^3$) and Se($4s^2 4p^4$), and we use the PBEsol form of the generalized gradient approximation (GGA) for the exchange-correlation functional. No Hubbard or Hund's exchange correction is added. We converge the Hellmann-Feynman forces to $10^{-5}$~eV/\AA{} using a plane-wave energy cutoff of 500~eV and a $11\times11\times11$ k-point mesh. The lattice constants of our fully relaxed structure, $\textbf{a}=4.13$~\AA{} and $\textbf{c}=28.6$~\AA{}, fit reasonably well to experimental values. Further computational details are provided in Supplementary Text S3.

To determine the single-mode anharmonicities and nonlinear phonon couplings, we compute the total energy of the system on a $11\times11\times11$ grid of atomic displacements along the eigenvectors of each of the phonon modes, $Q_{\text{E}^{a}_\text{u}}$, $Q_{\text{E}^{b}_\text{u}}$, $Q_{\text{E}^{a}_\text{g}}$ and $Q_{\text{E}^{b}_\text{g}}$ (Fig.~S2c,d), and fit the resulting potential energy landscape to the phonon potential energy, $V$. To control the displacement direction, we rotate the mode-effective charge vectors of the $\text{E}_\text{u}$ modes to align them with the $x$ and $y$ axes. The indices $a$ and $b$ correspond to this Cartesian system. This rotation mixes the eigenvectors, weighted by the rotation matrix elements, which are then used to generate the displacement grid. For the $\text{E}_\text{g}$ modes, we apply a different rotation such that the resulting Raman tensors correspond to Eq.~\ref{Raman_tensors}. To compute the Raman tensors, we determine the static dielectric function from DFT calculations using \textsc{vasp} and evaluate the tensors as derivatives of the dielectric function with respect to the Raman phonon amplitudes (see Eq.~\ref{eq:Raman_tensor}). We obtain a value of a Raman tensors' element $a=232$~\AA{}$^2/\sqrt{u}$, where $u$ is the atomic mass unit.

To compute the phonon angular momentum in Fig.~3, we evaluate $(0,0,L_z(t))=\boldsymbol{Q}(t)\times\dot{\boldsymbol{Q}}(t)$, with $\boldsymbol{Q}_{\text{E}_\text{u}}=(Q_{\text{E}^{a}_\text{u}},Q_{\text{E}^{b}_\text{u}},0)$ and $\boldsymbol{Q}_{\text{E}_\text{g}}=(Q_{\text{E}^{a}_\text{g}},Q_{\text{E}^{b}_\text{g}},0)$. The mode amplitudes for the phononic pathway are obtained by solving the coupled equations of motion (Eqs. \ref{eq:Q_1_time_IRS}-\ref{eq:Q_d_IRS}) using the full anharmonic potential $V$ of Eq.~\ref{eq:nonrotatedphononpotential}, while the photonic pathway is computed from Eqs.~\ref{eq:Q_c_SFE}-\ref{eq:Q_d_SFE}. In both cases we drive the system with the experimental THz field (Fig.~1e). Real-space atomic trajectories (visualized in Fig.~S3, Supplementary Movie~S3) are reconstructed by projecting the calculated amplitudes $\boldsymbol{Q}_{\text{E}_\text{u}}(t)$ and $\boldsymbol{Q}_{\text{E}_\text{g}}(t)$ onto the DFT eigenvectors (Fig.~S2c,d).

\subsection*{Angular Momentum Conservation}\label{Msubsec9}
The observed helicity reversal of the $\text{E}_\text{g}$ modes arises from angular momentum conservation between the $\text{E}_\text{u}$ and $\text{E}_\text{g}$ modes in the three-fold crystal lattice, which dictates the form of the driving force. Specifically, the phononic driving force originates from the symmetry-allowed coupling between the IR-active $\text{E}_\text{u}$ mode (transforming as a vector under the $\text{D}_{3\text{d}}$ point group) and the Raman-active $\text{E}_\text{g}$ mode (transforming as a symmetric second-rank tensor). Consequently, the driving force acquires the opposite helicity of the $\text{E}_\text{u}$ phonon generating it. This helicity reversal is shown in Fig.~4a-c, where Fig.~4a depicts the trajectory of the $\text{E}_\text{u}$ phonon, and Fig.~4b illustrates the resulting force acting on the $\text{E}_\text{g}$ mode, rotating in the opposite direction and thus carrying an inverted angular momentum. A detailed step-by-step modeling of the $\text{E}_\text{g}$ phonon dynamics, demonstrating the helicity flip, is provided in Extended Data Fig.~6.

The change of the helicity state becomes more evident when transforming from the Cartesian basis to a circular basis $(\hat{R},\hat{L})$. In the Cartesian basis, the driving force components are given by
\begin{equation}
F^{\text{ion}}_{\alpha} = c\,[Q^{\text{IR}^2}_x - Q^{\text{IR}^2}_y], \quad
F^{\text{ion}}_{\beta} = -2c\,Q^{\text{IR}}_x Q^{\text{IR}}_y.
\end{equation}
To express the force in a circular basis, we rewrite the phonon components and the corresponding driving force as
\begin{equation}
Q^{\text{IR}}_{|R\rangle} = \frac{1}{\sqrt{2}}[Q^{\text{IR}}_x + iQ^{\text{IR}}_y],\quad
Q^{\text{IR}}_{|L\rangle} = \frac{1}{\sqrt{2}}[Q^{\text{IR}}_x - iQ^{\text{IR}}_y],
\end{equation}
\begin{equation}
F^{\text{ion}}_{|R\rangle} = \frac{1}{\sqrt{2}}[F^{\text{ion}}_{\alpha} + iF^{\text{ion}}_{\beta}],\quad
F^{\text{ion}}_{|L\rangle} = \frac{1}{\sqrt{2}}[F^{\text{ion}}_{\alpha} - iF^{\text{ion}}_{\beta}].
\end{equation}
Here, $Q^{\text{IR}}_{|R\rangle}$ and $Q^{\text{IR}}_{|L\rangle}$ are the right- and left-handed circularly polarized states of the $\text{E}_\text{u}$ phonon, while $F^{\text{ion}}_{|R\rangle}$ and $F^{\text{ion}}_{|L\rangle}$ represent the phononic forces driving the corresponding right- and left-handed circular polarized states of the $\text{E}_\text{g}$ mode, respectively. In the circular basis, the driving force takes the form
\begin{equation}
F^{\text{ion}}_{|R\rangle}=c\,Q^{\text{IR}^2}_{|L\rangle}, \quad
F^{\text{ion}}_{|L\rangle}=c\,Q^{\text{IR}^2}_{|R\rangle}.
\end{equation}
In the circular basis, it becomes clear that the left-circularly polarized component of the $\text{E}_\text{u}$ phonon drives the right-circularly polarized $\text{E}_\text{g}$ phonon, and vice versa. This coupling leads to the observed helicity reversal, where the $\text{E}_\text{g}$ mode inherits the opposite helicity of the $\text{E}_\text{u}$ phonon.

\section*{Data availability}
All data used to generate the plots in this paper are publicly available via Zenodo (https://doi.org/10.5281/zenodo.19087102)~\cite{Zenodo2026}. Source data are provided with this paper.

\makeatletter
\renewcommand{\fnum@figure}{\textbf{Extended Data Fig. \thefigure}}
\setcounter{figure}{0} 
\makeatother

\begin{figure} 
	\centering
	\includegraphics[width=0.73\textwidth]{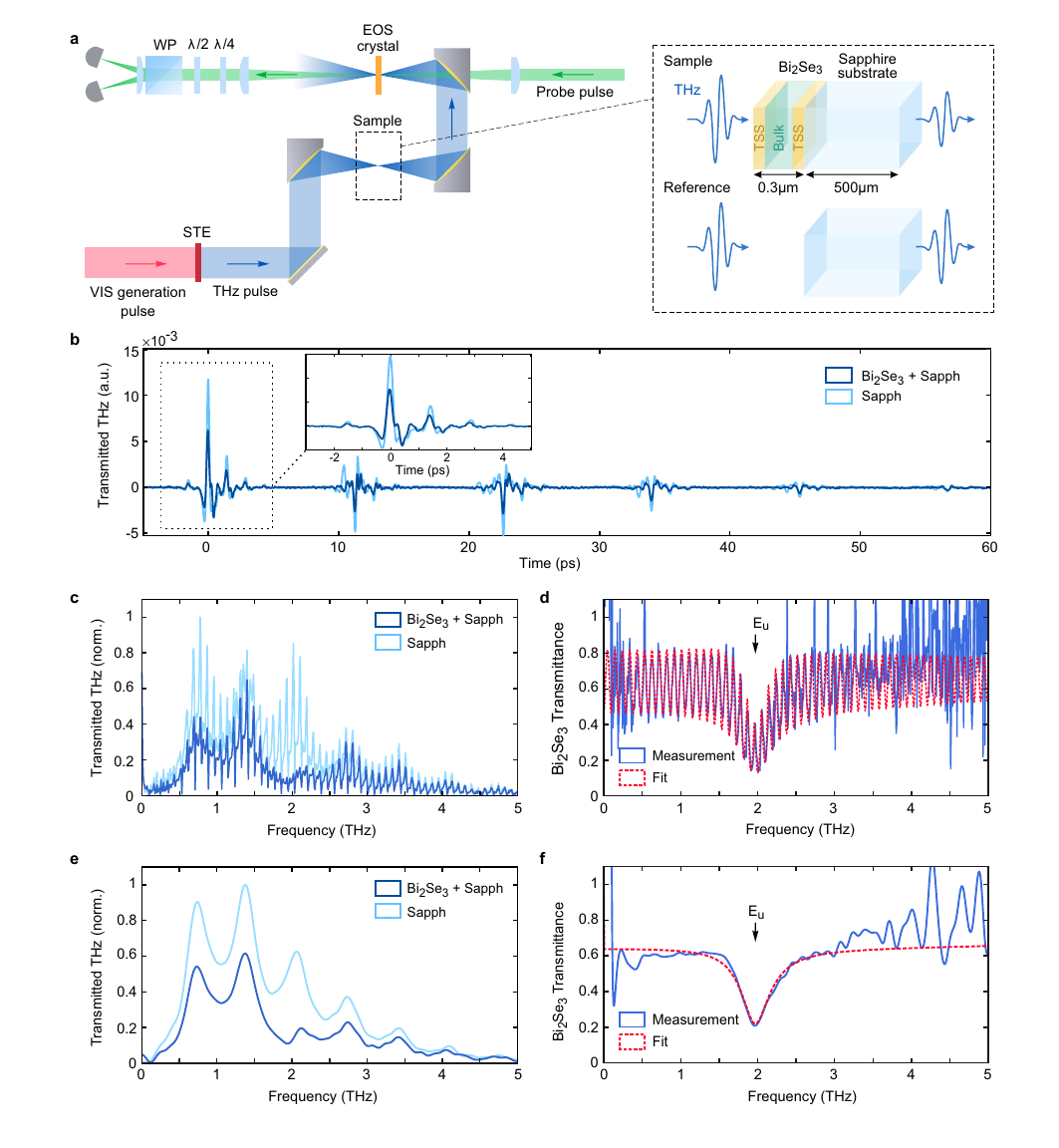}
	\caption{\textbf{Terahertz time-domain spectroscopy in $\text{Bi}_2\text{Se}_3$.}
		\textbf{a}, Schematic of the THz transmission setup. An ultrabroadband THz pulse from a large-area spintronic emitter is focused onto the sample position at the first focal point. The transmitted field is then refocused onto a 30~$\mu$m-thick GaP crystal for electro-optic detection. The sample position hosts either a 0.3~$\mu$m-thick $\text{Bi}_2\text{Se}_3$ film on a 500~$\mu$m sapphire substrate or a bare 500~$\mu$m-thick sapphire wafer for reference measurement. \textbf{b-c}, Time-domain signals of the sample (dark blue) and reference (light blue) THz transmission measurement, including their corresponding Fourier transforms. \textbf{d}, Division of the THz spectrum transmitted through the sample (panel \textbf{c}, dark blue) by the reference spectrum (panel \textbf{c}, light blue), revealing the THz transmittance of $\text{Bi}_2\text{Se}_3$ and the absorption feature of the $\text{E}_{\text{u}}$ phonon mode. The modeled $\text{Bi}_2\text{Se}_3$ transmittance function (red line) shows very good agreement. \textbf{e}, Fourier transforms of the fundamental transmitted pulse obtained by applying a 8~ps-time window to \textbf{b} (see inset). \textbf{f}, Corresponding ratio of transmitted THz spectra, as in \textbf{d}. Isolation of the fundamental transmitted pulses removes Fabry–Pérot oscillations from the transmission spectrum, enabling more precise modeling of the $\text{E}_{\text{u}}$ phonon feature.}
	\label{fig:extended_1} 
\end{figure}

\begin{figure} 
	\centering
	\includegraphics[width=0.95\textwidth]{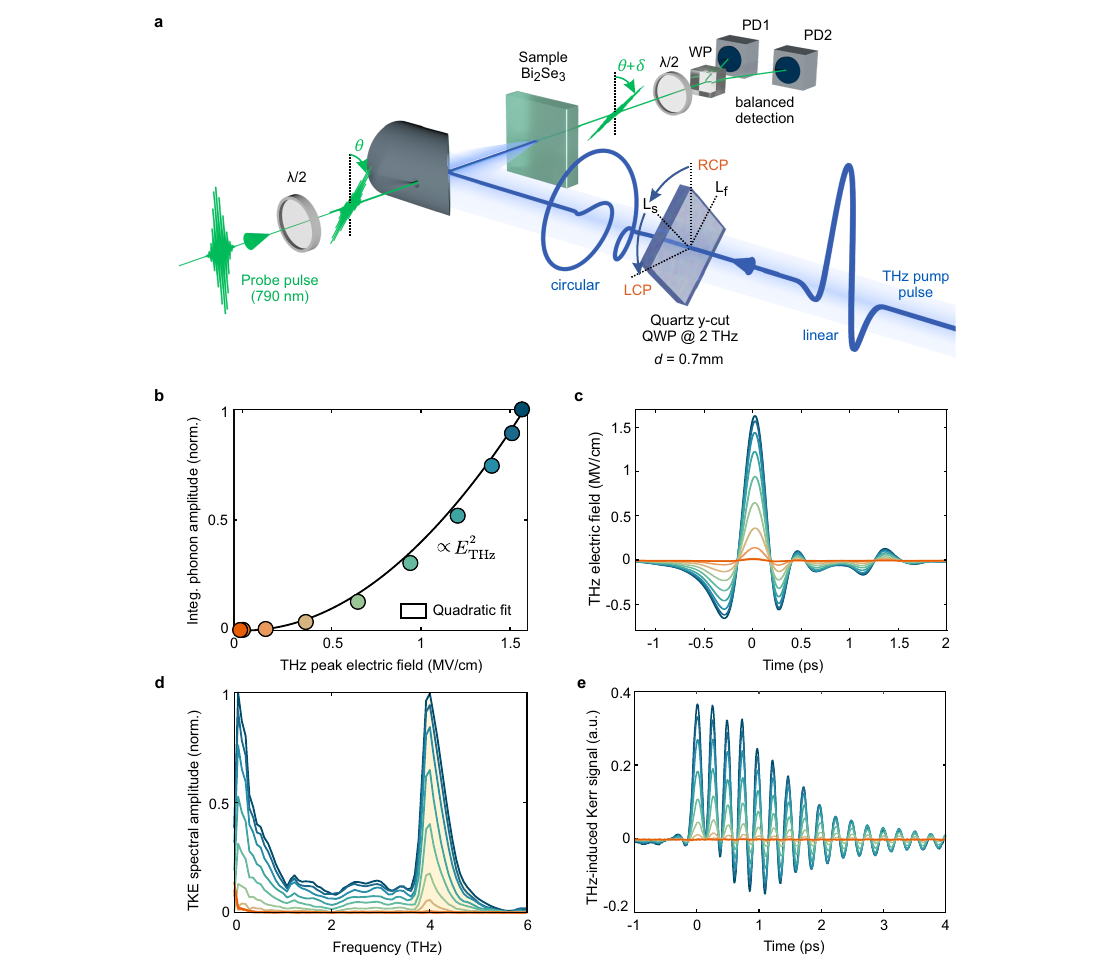}
	\caption{\textbf{Experimental TKE setup and THz field dependence.}
		\textbf{a}, Schematic of the experimental TKE setup. A THz pulse (blue), generated via optical rectification in $\text{LiNbO}_3$, passes through a 0.7~mm-thick y-cut quartz waveplate to acquire a defined ellipticity state before irradiating the sample at normal incidence. The excited Raman-active phonon is detected via transient birefringence, which induces a time-dependent rotation of the probe pulse (green) polarization measured by a balanced detection. The angle of the incident probe pulse’s linear polarization is controlled by a half-wave plate. \textbf{b-e}, THz fluence dependence of the Raman-active $\text{E}_{\text{g}}$ phonon. \textbf{b}, Integrated spectral amplitudes of the $\text{E}_{\text{g}}$ mode as a function of THz peak electric field (colored points) with a corresponding quadratic fit (black). \textbf{c}, Incident THz fields measured by electro-optic sampling in z-cut quartz. \textbf{d-e}, Corresponding TKE signals of $\text{Bi}_2\text{Se}_3$ in frequency and time domain, respectively. The shaded yellow area in \textbf{d} highlights the integration region used to determine the phonon amplitude.}
	\label{fig:extended_2} 
\end{figure}

\begin{figure} 
	\centering
	\includegraphics[width=\textwidth]{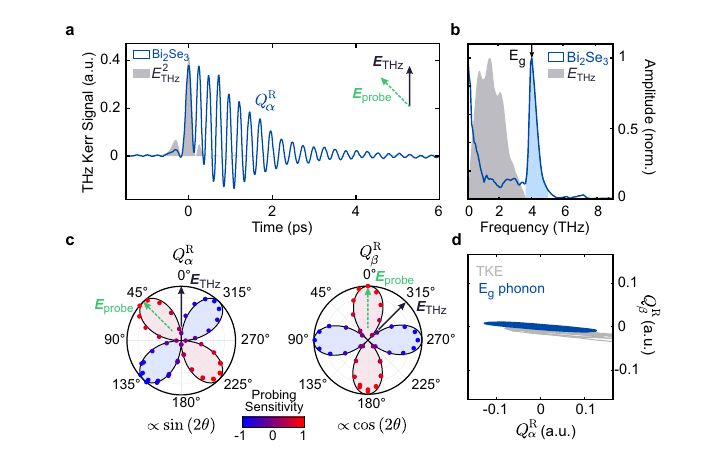} 
	\caption{\textbf{Probing phonon trajectories.}
		\textbf{a}, Measured TKE in $\text{Bi}_2\text{Se}_3$ (blue) excited by a linearly polarized THz pulse. The probe pulse is polarized at an angle of $45^{\circ}$ relative to the THz excitation pulse. The TKE signal around time zero includes an electronic contribution following $E^2_{\text{THz}}$ (gray shaded area). The oscillatory signal corresponds to an ionic contribution (phonon) to the TKE. \textbf{b}, Fourier transform of a, unveils the prominent $\text{E}_{\text{g}}$ phonon peak at 4~THz. \textbf{c}, Spectrally-integrated (panel \textbf{b}, blue shaded area) 4~THz coherent phonon amplitude (points) as a function of probe polarization angle (green arrow) relative to vertical and $45^{\circ}$-rotated THz excitation polarizations (black arrow), which drive orthogonal $\text{E}_{\text{g}}$ phonon polarizations. The theoretically calculated TKE amplitude and the positive/negative sign are shown by the black line and shaded red/blue areas, respectively. \textbf{d}, Phonon trajectory of the linearly polarized $\text{E}_{\text{g}}$ phonon measured with probe polarizations at $0^{\circ}$ ($\beta$-component) and $45^{\circ}$ ($\alpha$-component). The gray line represents the entire TKE signal, while the blue line shows the Fourier-filtered signal of the coherent phonon.}
	\label{fig:extended_3} 
\end{figure}

\begin{figure} 
	\centering
	\includegraphics[width=\textwidth]{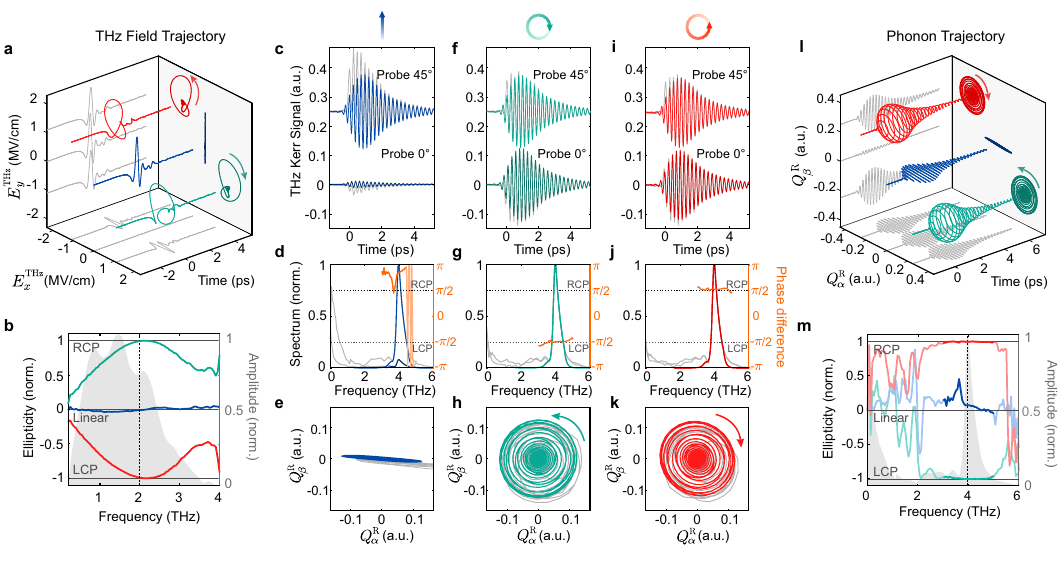} 
	\caption{\textbf{Lattice trajectory control by circular THz excitation in $\textrm{Bi}_2\textrm{Se}_3$.}
		\textbf{a}, Polarization-resolved EOS of the incident THz electric field transmitted through 700~$\mu$m-thick y-cut quartz at three azimuthal crystal orientations. \textbf{b}, Frequency-resolved helicity states of the THz fields in \textbf{a}. The shaded gray area represents the THz spectral amplitude. \textbf{c}-\textbf{k}, Measured $\textrm{E}_{\textrm{g}}$ trajectories under linear (blue), right-circular (green) and left-circular (red) THz excitation in $\textrm{Bi}_2\textrm{Se}_3$. \textbf{c}, Full TKE signals (gray) and Fourier-filtered phonon contributions (blue) measured with probe polarized at $0^{\circ}$ ($S_{0^\circ}$) and $45^\circ$ ($S_{45^\circ}$), vertically offset for clarity, corresponding to two orthogonal $\textrm{E}_{\textrm{g}}$ phonon components. \textbf{d}, Spectral amplitudes of signals in \textbf{c}, with the relative phase difference of the TKE signals $S_{45^\circ}-S_{0^\circ}$ (orange). \textbf{e}, Corresponding vectorial phonon trajectory. \textbf{f}-\textbf{k}, Same data set as \textbf{c}-\textbf{e}, but for right- and left-circular THz excitation. \textbf{l}, Coherent phonon trajectories for each THz helicity in \textbf{a}, extracted by Fourier-filtering the phonon contribution in the TKE signal. \textbf{m}, Frequency-resolved helicity of the $\textrm{E}_{\textrm{g}}$ phonon from \textbf{l}. The gray background represents the TKE spectrum, with the peak at 4~THz corresponding to the Raman-active $\textrm{E}_{\textrm{g}}$ phonon.}
	\label{fig:extended_4} 
\end{figure}

\begin{figure} 
	\centering
	\includegraphics[width=\textwidth]{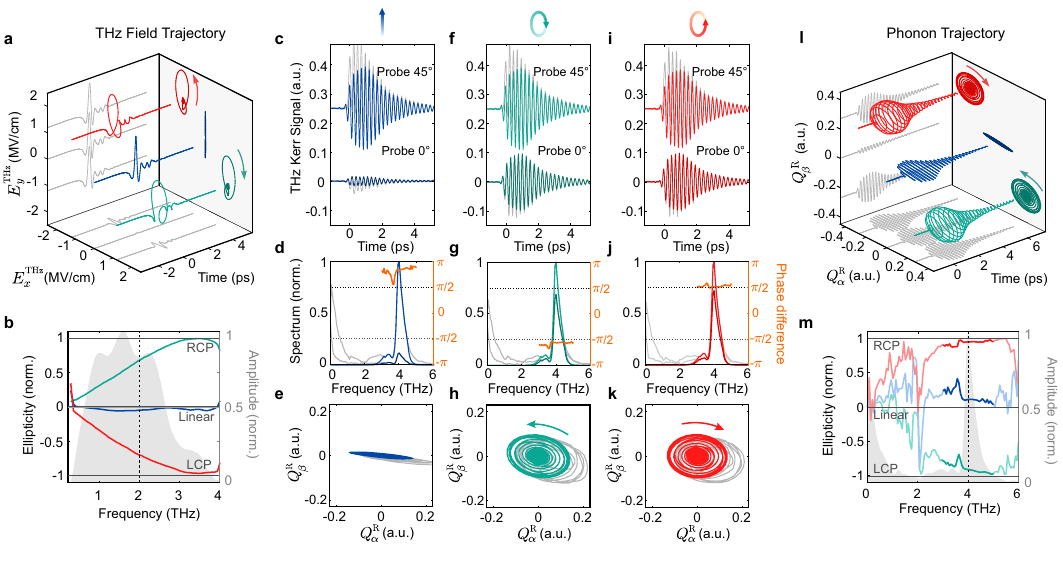} 
	\caption{\textbf{Lattice trajectory control by arbitrary elliptical THz excitation in $\textrm{Bi}_2\textrm{Se}_3$.}
		Same as Extended Data Fig.~\ref{fig:extended_4}, but for the incident THz electric field transmitted through 380~$\mu$m-thick y-cut quartz waveplate.}
	\label{fig:extended_5} 
\end{figure}

\begin{figure} 
	\centering
	\includegraphics[width=\textwidth]{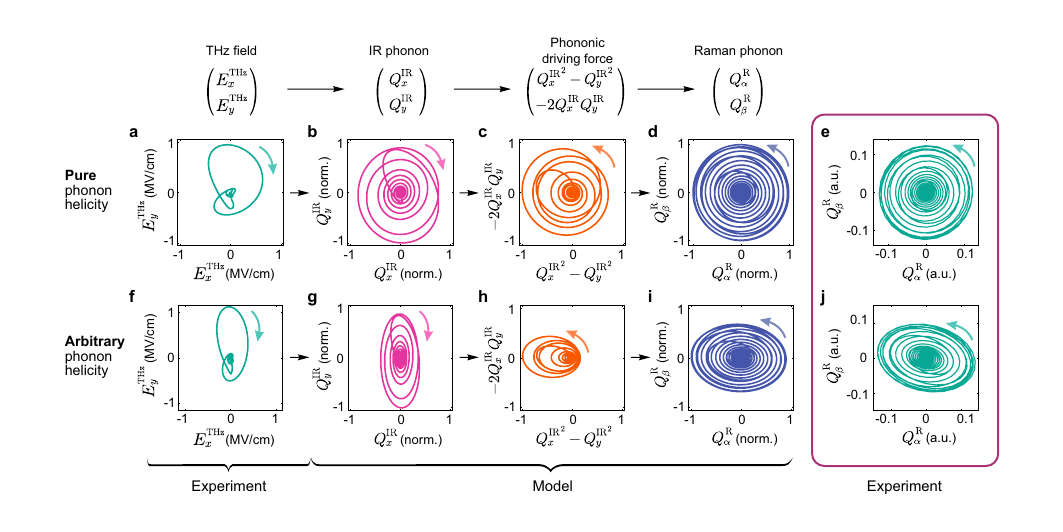} 

	\caption{\textbf{Modeling the phonon trajectory in a classical picture.}
		\textbf{a}-\textbf{d}, \textbf{f}-\textbf{i}, Step-by-step modeling of the $\textrm{E}_{\textrm{g}}$ phonon trajectory for purely circular (above) and arbitrary elliptical (below) THz excitation field. \textbf{a}, Vectorial trajectory of the right-handed circularly polarized THz electric field, measured by polarization-resolved EOS (Fig.~\ref{fig:main_2}a). The arrow indicates the direction of rotation. \textbf{b}, Corresponding IR-active $\textrm{E}_{\textrm{u}}$ phonon trajectory, calculated by solving the equation of motion with the driving force set by the experimental THz pulse $F_i^{\textrm{IR}} (t) \propto E_i^{\textrm{THz}}(t)-\partial V/\partial Q_i^{\text{IR}}, i\in\{x,y\}$. \textbf{c}, Phononic driving force acting on the Raman-active $\textrm{E}_{\textrm{g}}$ phonon, derived from the symmetry-allowed anharmonic coupling potential $V(\boldsymbol{Q}^{\text{IR}},\boldsymbol{Q}^{\textrm{R}})=c(Q_y^{\text{IR}^2}-Q_x^{\text{IR}^2})Q_{\alpha}^{\textrm{R}}+2cQ_x^{\text{IR}}Q_y^{\text{IR}} Q_{\beta}^{\textrm{R}}$ between $\textrm{E}_{\textrm{u}}$ and $\textrm{E}_{\textrm{g}}$ modes as $F_i^{\textrm{R}} (t) \propto -\partial V/\partial Q_i^{\text{R}}, i\in\{\alpha,\beta\}$. The driving force exhibits opposite helicity to the IR phonon in \textbf{b}. \textbf{d}, Calculated trajectory of Raman-active $\textrm{E}_{\textrm{g}}$ phonon. \textbf{e}, Experimentally-retrieved $\textrm{E}_{\textrm{g}}$ phonon trajectory, repeated from Fig.~\ref{fig:main_2}e. \textbf{f}-\textbf{j}, Same modeling procedure for arbitrary elliptically polarized THz excitation (Extended Data Fig.~\ref{fig:extended_5}a), with comparison to the experiment (Extended Data Fig.~\ref{fig:extended_5}h).}
	\label{fig:extended_6} 
\end{figure}




\newpage


\makeatletter
\renewcommand{\fnum@figure}{\textbf{Fig. \thefigure}}
\makeatother
\renewcommand{\thefigure}{S\arabic{figure}}
\renewcommand{\thetable}{S\arabic{table}}
\renewcommand{\theequation}{S\arabic{equation}}
\renewcommand{\thepage}{S\arabic{page}}
\setcounter{figure}{0}
\setcounter{table}{0}
\setcounter{equation}{0}
\setcounter{page}{1} 


\newpage
\begin{center}
\section*{Supplementary Information for\\ \scititle}

    Olga~Minakova,
    Carolina~Paiva,
    Maximilian~Frenzel,
    Michael~S.~Spencer,
    Joanna~M.~Urban,
    Christoph~Ringkamp,
    Martin~Wolf,
    Gregor~Mussler,
    Dominik~M.~Juraschek,
    Sebastian~F.~Maehrlein$^{\ast}$\\
\small$^\ast$Corresponding author. Email: s.maehrlein@hzdr.de\\
\end{center}

\subsubsection*{This PDF file includes:}
Supplementary Text\\
Table S1\\
Figures S1 to S8\\
Captions for Movies S1 to S3\\

\subsubsection*{Other Supplementary Information for this manuscript:}
Movies S1 to S3\\

\newpage
\section*{Table of Contents}

\noindent\textbf{Supplementary Text}

S1.\ Symmetry analysis and basis convention for $\text{E}_\text{u}$ and $\text{E}_\text{g}$ phonon modes
\dotfill \pageref{sec:S1}

S2.\ Theory of phonon angular momentum upconversion in first-principles calculations
\dotfill \pageref{sec:S2}

S3.\ Details of first-principle calculations
\dotfill \pageref{sec:S3}

S4.\ Analytical model for THz transmission measurement of Bi$_2$Se$_3$
\dotfill \pageref{sec:S4}

References
\dotfill \pageref{ref:SI}

\vspace{0.5em}
\noindent\textbf{Supplementary Tables}

Table S1.\ Summary of calculated phonon properties and anharmonic coefficients
\dotfill \pageref{table}

\vspace{0.5em}
\noindent\textbf{Supplementary Figures}

Figure S1.\ Vibrational modes of the $\text{C}_\text{3v}$ and $\text{D}_\text{3d}$ model systems
\dotfill \pageref{fig:S1}

Figure S2.\ Symmetry-adapted eigenvectors and basis handedness
\dotfill \pageref{fig:S2}

Figure S3.\ Computed real-space atomic trajectories
\dotfill \pageref{fig:S3}

Figure S4.\ Multilayer thin-film interference model
\dotfill \pageref{fig:S4}

Figure S5.\ X-ray reflectivity (XRR)
\dotfill \pageref{fig:S5}

Figure S6.\ X-ray diffraction (XRD)
\dotfill \pageref{fig:S6}

Figure S7.\ Characterization of y-cut quartz waveplates by THz electro-optic sampling
\dotfill \pageref{fig:S7}

Figure S8.\ Absence of transient magnetic contribution in the Kerr response
\dotfill \pageref{fig:S8}

\vspace{0.5em}
\noindent\textbf{Supplementary Movies (captions)}

Movie S1.\ Experimental THz electric field and measured $\text{E}_\text{g}$ phonon trajectories
\dotfill \pageref{mov:S1}

Movie S2.\ Simulated coupled $\text{E}_\text{u}$ and $\text{E}_\text{g}$ phonon trajectories
\dotfill \pageref{mov:S2}

Movie S3.\ Time evolution of phonon helicity reversal from \textit{ab initio} DFT
\dotfill \pageref{mov:S3}


\newpage

\subsection*{Supplementary Text}

\subsubsection*{ S1. Symmetry analysis and basis convention for $\text{E}_\text{u}$ and $\text{E}_\text{g}$ phonon modes}\label{sec:S1}

A proper analysis of the coupling symmetry between degenerate phonon modes, particularly their ellipticity states (RCP/LCP), requires an internally consistent basis convention. This is essential for two-dimensional degenerate modes, as their choice of basis is not unique. In the $\text{D}_\text{3d}$ symmetry, the $\text{E}_\text{g}$ mode is commonly represented by the pair of basis functions $x^2-y^2$ and $2xy$, although there is a freedom in their relative sign, e.g. $(x^2-y^2,2xy)$ vs. $(x^2-y^2,-2xy)$. While these choices are mathematically equivalent, the sign of the $\pm2xy$ component determines the handedness of the $\text{E}_\text{g}$ basis (right- or left-handed), which should be aligned with the handedness of the $(x,y)$ basis of the $\text{E}_\text{u}$ mode for a clear analysis.

The specific form of the anharmonic coupling potential used in the main text, $V \propto [(Q^{\text{IR}}_x)^2-(Q^{\text{IR}}_y)^2]Q^{\text{R}}_\alpha -2Q^{\text{IR}}_xQ^{\text{IR}}_yQ^{\text{R}}_\beta$, follows directly from the basis convention defined for the atomic displacements in $\text{E}_\text{u}$ and $\text{E}_\text{g}$ modes. The sign of the second term is critical; a positive sign (e.g. $+2Q^{\text{IR}}_xQ^{\text{IR}}_yQ^{\text{R}}_\beta$) yields a driving force $\mathbf{F} = (-\partial V/\partial Q^{\text{R}}_\alpha, -\partial V/\partial Q^{\text{R}}_\beta)\propto (Q^{\text{IR\;2}}_x-Q^{\text{IR}\;2}_y, +2Q^{\text{IR}}_xQ^{\text{IR}}_y)$ that rotates in the same direction as the $\text{E}_\text{u}$ mode $Q^{\text{IR}}$, obscuring the helicity reversal observed experimentally and reproduced by DFT. This discrepancy arises from an inconsistent definition of the coordinate systems of IR and Raman modes, where these bases possess opposite handedness. In this section, we demonstrate that preserving a consistent, right-handed coordinate system for both the $\text{E}_\text{u}$ $(x,y)$ mode and the $\text{E}_\text{g}$ mode requires the representative $\text{E}_\text{g}$ basis to transform as $(x^2 - y^2,-2xy)$. This choice establishes the negative sign in the coupling potential, thereby aligning the bases and correctly capturing the counter-rotating driving force reproduced also by DFT calculations.

To resolve the basis ambiguity outlined above, we determine the vibrational eigenvectors of $\text{E}_\text{u}$ and $\text{E}_\text{g}$ modes for a model system obeying the $\text{D}_\text{3d}$ symmetry, as in $\text{Bi}_\text{2}\text{Se}_\text{3}$. We restrict the analysis to in-plane displacements, representing the model as two stacked, inverted triangles at $+z$ and $-z$, forming a $\text{D}_\text{3d}$ structure with an inversion center at the origin. The vibrational $\text{E}_\text{u}$ and $\text{E}_\text{g}$ eigenvectors of this 6-atom system can be built from the E-symmetry mode of a single $\text{C}_\text{3v}$ triangle (Fig.~4d), extending it to the second triangle based on the mode parity under the inversion $i$. The $\text{E}_\text{u}$ mode is odd under inversion, so a displacement $(x,y)$ on the top triangle is matched by an identical $(x,y)$ displacement on its inversion partner. In contrast, the $\text{E}_\text{g}$ mode is even, therefore a $(x,y)$ displacement on the top atom corresponds to an inverted $(-x,-y)$ displacement on its partner (see Figure~S1).

\begin{figure} 
	\centering
	\includegraphics[width=0.9\textwidth]{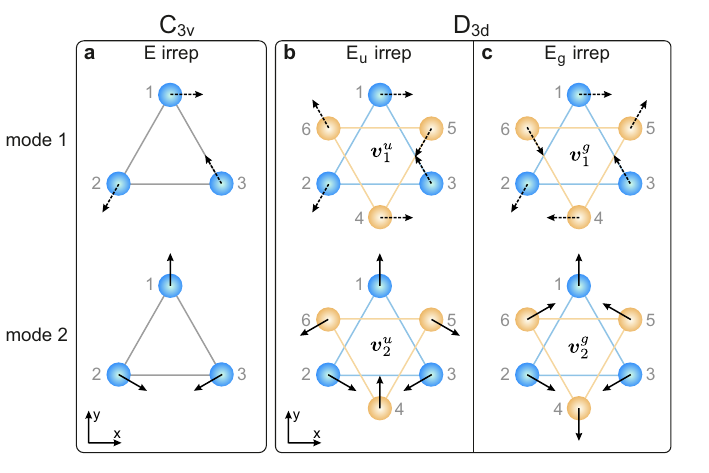} 
    
	\caption{\textbf{Vibrational modes of the $\textrm{C}_\textrm{3v}$ and $\textrm{D}_\textrm{3d}$ model systems.}
		\textbf{a}, The doubly degenerate E-symmetry modes of a single triangle $\textrm{C}_\textrm{3v}$. \textbf{b}-\textbf{c}, The corresponding $\text{E}_\text{u}$ and $\text{E}_\text{g}$ eigenmodes for the $\textrm{D}_\textrm{3d}$ system. The $\text{E}_\text{u}$ modes (\textbf{b}) exhibit odd parity under inversion, while the $\text{E}_\text{g}$ modes (\textbf{c}) have even parity. These eigenvectors represent the physical basis obtained using the projection operator method.}
	\label{fig:S1} 
\end{figure}

We directly determine these vibrational modes using the projection operator method, a standard technique from group theory. To do this, we describe the system in a 12-dimensional basis representing the $(x,y)$ displacements of the six atoms, $\textbf{v} = [x_1, y_1, x_2, y_2,\dots,x_6, y_6]$. Modes belonging to a given irreducible representation (irrep) $\Gamma_\text{j}$ are those that transform according to that irrep, and can be obtained by applying the projection operator $P^{\Gamma_\text{j}} = (l_\text{j}/h)\sum_{R} \chi^{(\text{j})}(R)^* D(R)$ to a given vector in the space, extracting the $\Gamma_\text{j}$ component of a vector in the representation space. Here, $l_\text{j}$ is the dimension of $\Gamma_\text{j}$, $h$ is the order of the group, $\chi^{(\text{j})}(R)$ is its character for a symmetry operation $R$, and $D(R)$ is the $12\times12$ matrix representation of $R$, incorporating both the spatial action on $(x, y)$ and the atomic permutation. Diagonalizing $P^{\text{E}\text{u}}$ and $P^{\text{E}\text{g}}$ yields their eigenvectors $v^{u/g}_{i}$ with eigenvalue 1, corresponding to the in-plane vibrational modes of the six atoms (Figure~S1b,c). These eigenvectors form an arbitrary orthogonal basis spanning each subspace. To straightforwardly connect with mode helicity, we map them onto standard basis functions: the vector basis $(x,y)$ for the $\text{E}_\text{u}$ mode and the quadratic basis $(x^2-y^2, \pm 2xy)$ for the $\text{E}_\text{g}$ mode.

To connect the atomic vibrational modes to the mathematical basis, we first analyze their behavior under a symmetry operation that distinguishes each basis component. A convenient choice is the $C_2(x)$ rotation (a $\pi$ rotation about the x-axis). Under $C_2(x)$, the atoms permute as $1\leftrightarrow4$, $2\leftrightarrow6$, $3\leftrightarrow5$, while the spatial coordinates transform as $x\rightarrow x$, $y\rightarrow -y$. As a result, $x$ and $x^2-y^2$ basis functions transform symmetrically (eigenvalue +1), whereas $y$ and $\pm 2xy$ transform antisymmetrically (eigenvalue -1).
The representation of the $C_2(x)$ symmetry operation in the 12D space has the form

\begin{equation}
    D(C_2) = \begin{pmatrix}
\mathbf{0} & \mathbf{0} & \mathbf{0} & \boldsymbol{C}_2 & \mathbf{0} & \mathbf{0} \\
\mathbf{0} & \mathbf{0} & \mathbf{0} & \mathbf{0} & \mathbf{0} & \boldsymbol{C}_2 \\
\mathbf{0} & \mathbf{0} & \mathbf{0} & \mathbf{0} & \boldsymbol{C}_2 & \mathbf{0} \\
\boldsymbol{C}_2 & \mathbf{0} & \mathbf{0} & \mathbf{0} & \mathbf{0} & \mathbf{0} \\
\mathbf{0} & \mathbf{0} & \boldsymbol{C}_2 & \mathbf{0} & \mathbf{0} & \mathbf{0} \\
\mathbf{0} & \boldsymbol{C}_2 & \mathbf{0} & \mathbf{0} & \mathbf{0} & \mathbf{0}
\end{pmatrix}, 
\quad \boldsymbol{C}_2=\begin{pmatrix}
1 & 0 \\ 0 & -1
\end{pmatrix},
\; \mathbf{0}=\begin{pmatrix}
0 & 0 \\ 0 & 0
\end{pmatrix}.
\end{equation}
Here, $\boldsymbol{C}_2$ denotes the $2\times2$ rotation matrix of $C_2(x)$, $\mathbf{0}$ is the $2\times2$ zero matrix. The block structure encodes the atomic permutation combined with the vector transformation $\boldsymbol{C}_2$.

When this operator acts on the $\text{E}_\text{u}$ and $\text{E}_\text{g}$ eigenvectors from Figure~S1b,c, the result is
\[
\begin{array}{lclcl}
\multicolumn{2}{c}{{\text{E}_\text{u}:}} & \quad & \multicolumn{2}{c}{{\text{E}_\text{g}}:} \\[0.5em]
D(C_2)\,\mathit{v}^{\,u}_1 = +1 \cdot \mathit{v}^{\,u}_1 & 
\textcolor{purple!50}{\rightarrow \mathit{v}^{\,u}_1 \sim x} & \quad &
D(C_2)\,\mathit{v}^{\,g}_1 = -1 \cdot \mathit{v}^{\,g}_1 & 
\textcolor{blue!50}{\rightarrow \mathit{v}^{\,g}_1 \sim 2xy} \\[0.5em]
D(C_2)\,\mathit{v}^{\,u}_2 = -1 \cdot \mathit{v}^{\,u}_2 & 
\textcolor{blue!50}{\rightarrow \mathit{v}^{\,u}_2 \sim y} & \quad &
D(C_2)\,\mathit{v}^{\,g}_2 = +1 \cdot \mathit{v}^{\,g}_2 & 
\textcolor{purple!50}{\rightarrow \mathit{v}^{\,g}_2 \sim x^2-y^2}
\end{array}
\]

The $C_2(x)$ analysis assigns each eigenvector $v^{u/g}_i$ to a symmetric ($x$,\;$x^2-y^2$) or antisymmetric ($y$,\;$2xy$) component, leaving their relative sign, and thus the handedness of the coordinate systems, yet undetermined. To resolve this and establish a consistent definition of angular momentum for both modes, we examine their transformation under the $C_3$ rotation.

We apply $D(C_3)$ (representation of $2\pi/3$ rotation about the z-axis, $C_3$) to both $\text{E}_\text{u}$ $(v^{u}_1,\;v^{u}_2)$ and $\text{E}_\text{g}$ $(v^{g}_2,\;v^{g}_1)$ pairs of eigenvectors. By viewing the outcome of this basis transformation, we will identify a basis convention, where the right- and left-handed polarization is defined the same way for $\text{E}_\text{u}$ as for $\text{E}_\text{g}$ modes. The $D(C_3)$ matrix is given by

\begin{equation}
    D(C_3) = \begin{pmatrix}
\mathbf{0} & \mathbf{0} & \boldsymbol{C}_3 & \mathbf{0} & \mathbf{0} & \mathbf{0} \\
\boldsymbol{C}_3 & \mathbf{0} & \mathbf{0} & \mathbf{0} & \mathbf{0} & \mathbf{0} \\
\mathbf{0} & \boldsymbol{C}_3 & \mathbf{0} & \mathbf{0} & \mathbf{0} & \mathbf{0} \\
\mathbf{0} & \mathbf{0} & \mathbf{0} & \mathbf{0} & \mathbf{0} & \boldsymbol{C}_3 \\
\mathbf{0} & \mathbf{0} & \mathbf{0} &\boldsymbol{C}_3 & \mathbf{0} & \mathbf{0} \\
\mathbf{0} & \mathbf{0} & \mathbf{0} & \mathbf{0} & \boldsymbol{C}_3 & \mathbf{0}
\end{pmatrix}, 
\quad \boldsymbol{C}_3=\begin{pmatrix}
\cos{(2\pi/3)} & -\sin{(2\pi/3)} \\ \sin{(2\pi/3)} & \cos{(2\pi/3)}
\end{pmatrix},
\; \mathbf{0}=\begin{pmatrix}
0 & 0 \\ 0 & 0
\end{pmatrix}.
\end{equation}

The action of $D(C_3)$ on $\text{E}_\text{u}$ eigenvectors $(v^{u}_1,\;v^{u}_2)$ yields

\begin{equation}
D(C_3)\,\mathit{v}^{\,u}_1 \equiv \cos\!\left(\frac{2\pi}{3}\right) \mathit{v}^{\,u}_1 - \sin\!\left(\frac{2\pi}{3}\right) \mathit{v}^{\,u}_2,
\end{equation}
\begin{equation}
D(C_3)\,\mathit{v}^{\,u}_2 \equiv \sin\!\left(\frac{2\pi}{3}\right) \mathit{v}^{\,u}_1 + \cos\!\left(\frac{2\pi}{3}\right) \mathit{v}^{\,u}_2.
\end{equation}

While $D(C_3)$ acts in the full 12-dimensional displacement space, its action within the 2-dimensional $\text{E}_\text{u}$ subspace reduces to the $+2\pi/3$ rotation of the basis vectors $(v^{u}_1,\;v^{u}_2)$. This confirms that the subspace $(v^{u}_1,\;v^{u}_2)$ transforms as the vector coordinates $(x, y)$, thereby forming a 'right-handed' basis for the $\text{E}_\text{u}$ mode. This interpretation is supported by inspecting the atomic displacements in Fig.~S1b, where  the eigenvector pair $(v^{u}_1,\;v^{u}_2)$ forms a right-handed orthogonal basis on each atomic site. Consequently, we establish the following mapping
\[
v^{u}_1 \leftrightarrow x, \quad v^{u}_2 \leftrightarrow y.
\]

To analyze the $\text{E}_\text{g}$ eigenvectors $(v^{g}_2,\;v^{g}_1)$, we first recall that the associated quadratic basis functions $x^2-y^2$ and $2xy$ rotate by $2\theta$, unlike vectors, which rotate by $\theta$. Specifically, under a coordinate rotation of angle $+\theta$, the basis pair transforms as
\[
(x^2-y^2, \pm 2xy) \rightarrow R(\pm 2\theta)(x^2-y^2, \pm 2xy),
\]
where the sign of the second component determines the direction of the induced rotation $R(\pm 2\theta)$. Applying the $D(C_3)$ to the specific eigenvectors yields a rotation of $+4\pi/3$ within the subspace $(v^{g}_2,\;v^{g}_1)$

\begin{figure} [t!] 
	\centering
	\includegraphics[width=0.7\textwidth]{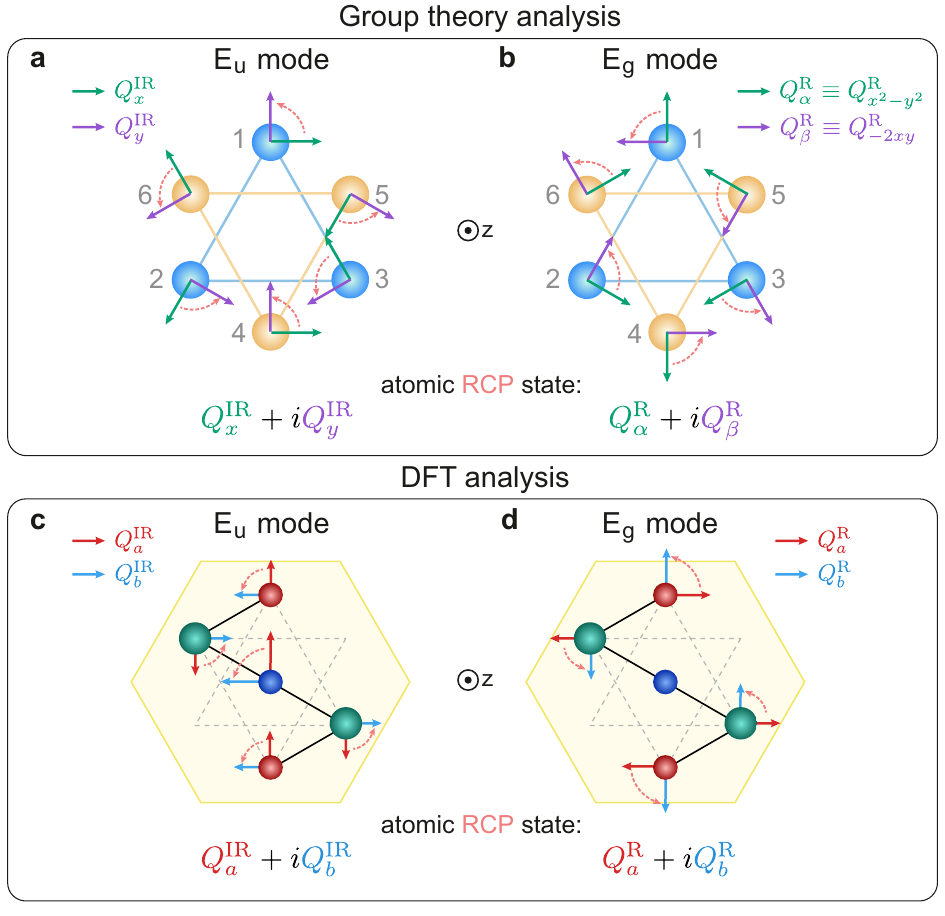} 
	\caption{\textbf{ Symmetry-adapted eigenvectors and basis handedness.}
		\textbf{a}-\textbf{b}, Model eigenvectors for $\text{E}_\text{u}$ and $\text{E}_\text{g}$ modes derived by diagonalizing the projection operators $P^{\text{E}\text{u}}$ and $P^{\text{E}\text{g}}$. The $(Q^\text{IR}_x, Q^\text{IR}_y)$ eigenvector pair transforms as $(x,y)$, while $(Q^\text{R}_{\alpha}, Q^\text{R}_{\beta})$ transforms as $(x^2-y^2,-2xy)$, both forming right-handed orthogonal bases at each atomic site with the z-axis aligned in the same direction. This basis convention enforces the coupling potential to be as Eq.\ref{eq:conv_coupling} and correctly captures the helicity reversal. \textbf{c}-\textbf{d}, Eigenvectors computed from \textit{ab-initio} DFT. These modes exhibit the same local basis orientation as the model, consistent with the derived potential $V$.}
	\label{fig:S2} 
\end{figure}

\begin{equation}
D(C_3)\,\mathit{v}^{\,g}_2 \equiv \cos\!\left(2\cdot\frac{2\pi}{3}\right) \mathit{v}^{\,g}_2 - \sin\!\left(2\cdot \frac{2\pi}{3}\right) \mathit{v}^{\,g}_1,
\end{equation}
\begin{equation}
D(C_3)\,\mathit{v}^{\,g}_1 \equiv \sin\!\left(2\cdot\frac{2\pi}{3}\right) \mathit{v}^{\,g}_2 + \cos\!\left(2\cdot\frac{2\pi}{3}\right) \mathit{v}^{\,g}_1.
\end{equation}

This rotation of $+2\cdot2\pi/3$ identifies the pair with the quadratic basis $(x^2-y^2,+2xy)$. The result is consistent with the atomic displacements in Fig.~S1c, where the pair $(v^{g}_2,\;v^{g}_1)$ forms a left-handed orthogonal basis on each atomic site, $v^{g}_2-iv^{g}_1$, relative to the right-handed $\text{E}_\text{u}$ $(v^{u}_1,\;v^{u}_2)$ reference. In a three-fold symmetric system, the phase accumulation of $+4\pi/3$ is equivalent to $-2\pi/3$, representing a counter-rotation relative to the $\text{E}_\text{u}$ reference frame. To enforce a consistent convention, where both bases accumulate the same phase under rotation, we select the basis $(v^{g}_2,\;-v^{g}_1)$. This choice maps to  $(x^2-y^2,-2xy)$ and inverts the rotation in the subspace to $-4\pi/3\equiv +2\pi/3$.

\begin{figure} [t!] 
	\centering
	\includegraphics[width=\textwidth]{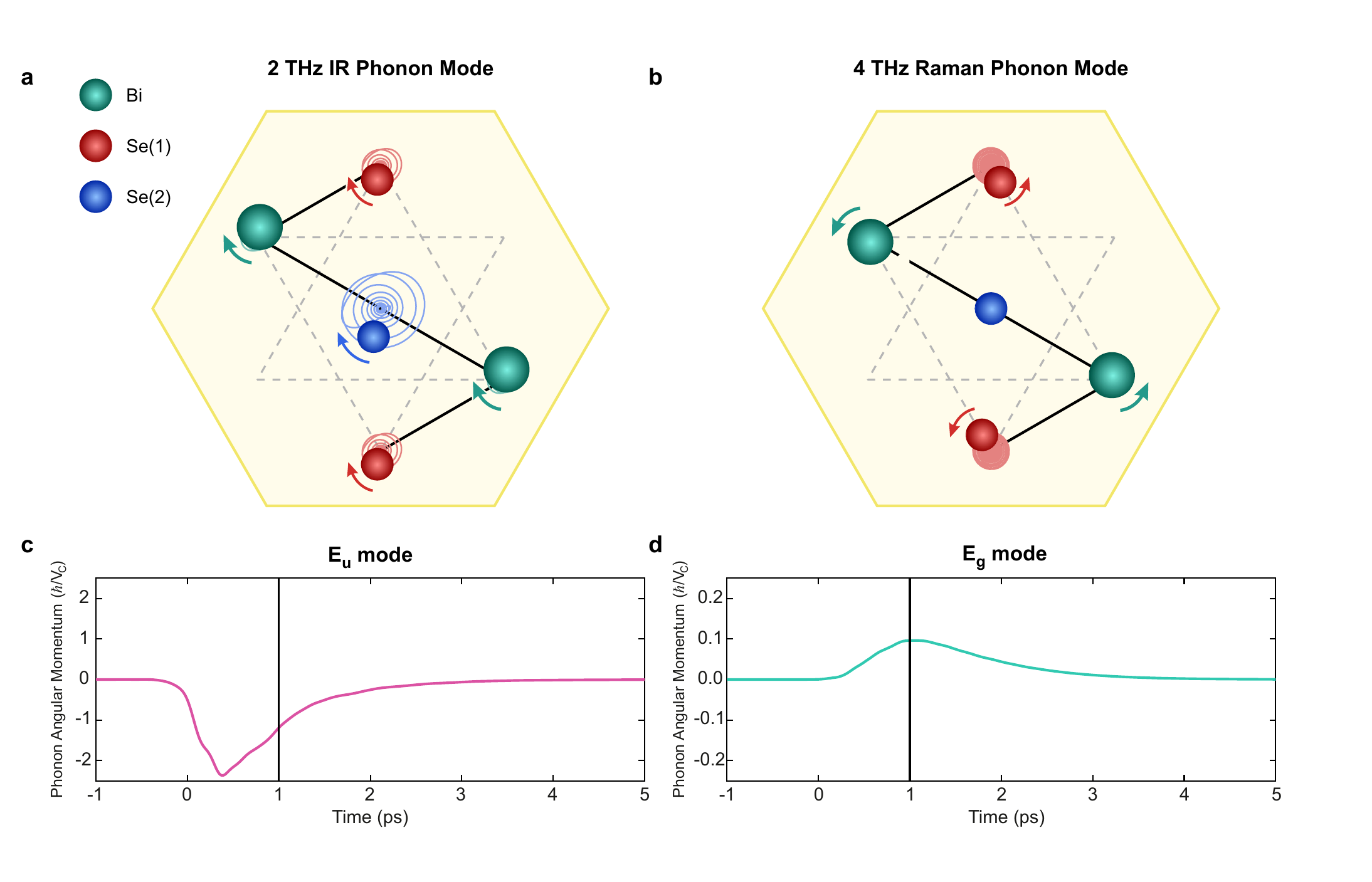} 
	\caption{\textbf{Computed real-space atomic trajectories.}
		\textbf{a}-\textbf{b}, Time-evolution of atomic displacements within the unit cell of $\text{Bi}_2\text{Se}_3$, obtained by solving the system of coupled equations of motion (Eqs.\ref{eq:Q_1_time_IRS}-\ref{eq:Q_d_IRS}) with the full form of the coupling potential $V$ (Eq.\ref{eq:nonrotatedphononpotential}). The solution is projected onto the \textit{ab-initio} eigenvectors shown in Fig.S2c,d. Phonon amplitudes are normalized relative to each other for clearer visibility. The resulting paths visualize the predicted helicity reversal between the driven $\text{E}_\text{u}$ and the coupled $\text{E}_\text{g}$ modes (see Supplementary Movie S3). \textbf{c}-\textbf{d}, Quantitative \textit{ab-initio} modeling of angular momentum dynamics of $\text{E}_\text{u}$ and $\text{E}_\text{g}$ modes (see Fig.3)}
	\label{fig:S3} 
\end{figure}

In conclusion, to perform a classical analysis of the symmetry of anharmonic coupling between $\text{E}_\text{u}$ and $\text{E}_\text{g}$ phonon modes, the potential must be defined to preserve the same handedness for both phonon bases
\begin{equation}
V \propto c[(Q^{\text{IR}}_x)^2-(Q^{\text{IR}}_y)^2]Q^{\text{R}}_\alpha -2cQ^{\text{IR}}_xQ^{\text{IR}}_yQ^{\text{R}}_\beta.  
\label{eq:conv_coupling}
\end{equation}
This ensures that the circular coordinates, $Q^{\text{R}}_{\alpha}\pm iQ^{\text{R}}_{\beta}$ and $Q^{\text{IR}}_x \pm iQ^{\text{IR}}_y$ describe rotations in the same direction. For the \textit{ab-initio} calculations, however, such basis alignment is unnecessary, as the coupling is evaluated directly in the eigenvector basis obtained from DFT. The DFT results independently demonstrate the helicity reversal. To illustrate this, we compute the real-space atomic trajectories by solving the coupled equations of motion (Eqs. \ref{eq:Q_1_time_IRS}-\ref{eq:Q_d_IRS}) with the full coupling potential (Eq.\ref{eq:nonrotatedphononpotential}), using the \textit{ab-initio} eigenvectors of Fig.~S2c,d. The resulting time-dependent displacements are visualized in Fig.~S3 and Supplementary Movie S3.


\subsubsection*{S2. Theory of phonon angular momentum upconversion in first-principle calculations}\label{sec:S2}
The equations of motion describing the coherent driving of phonon modes can be expressed as \cite{Juraschek2018}
\begin{equation}
 \ddot{Q}_{\alpha} + \gamma_{\alpha} \dot{Q}_{\alpha} + \partial_{Q_{\alpha}} V = \sum_{i} Z_{i,\alpha}E_{i} + \epsilon_{0}\sum_{ij} R_{ij,\alpha}E_{i}E_{j}, 
\label{eq:general_equation_motion}
\end{equation}
where $Q_{\alpha}$ is the phonon amplitude, $\gamma_{\alpha}$ is the phonon linewidth, and $V$ is the phonon potential energy. The index $\alpha=\{\text{E}^{a}_\text{u},\text{E}^{b}_\text{u},\text{E}^{a}_\text{g},\text{E}^{b}_\text{g}\}$ denotes the two orthogonal branches of each of the doubly degenerate $\text{E}_\text{u}$ and $\text{E}_\text{g}$ modes. We note here that this \textit{ab-initio} theoretical treatment, which is referenced to the crystalline axes (a,b), is connected to the experimental (i.e., labframe coordinates - x,y) description by the relative rotational invariance of the forces, which follows from the rotational symmetry of the system. The right-hand side of the equation contains the light-matter interactions. The first term describes infrared absorption and contains the mode effective charge, $\boldsymbol{Z}_{\alpha}$, given by
\begin{equation}
 \boldsymbol{Z}_{\alpha}=\sum_n Z^{*}_{n} \frac{\boldsymbol{q}_{n,\alpha}}{\sqrt{M_n}}, 
\end{equation}
where $Z^{*}_{n}$ is the Born effective charge tensor of atom $n$, $\boldsymbol{q}_{n,\alpha}$ is its phonon eigenvector, and $M_n$ its atomic mass, and the sum runs over all atoms in the unit cell. The second term describes Raman scattering  and contains the Raman tensors, $R_{ij}$, given by
\begin{equation}
R_{ij,\alpha}=V_{c}\frac{\partial \mathbf{\epsilon}_{ij}}{\partial Q_{\alpha}},
\label{eq:Raman_tensor}
\end{equation}
where $\epsilon_{ij}$ is the dielectric function, $V_{c}$ the volume of the unit cell and $i,j$ denote the spatial coordinates. $\boldsymbol{E}(t)$ represents the electric field component of the driving THz pulse, polarized in the $ab$ plane of the crystal according to the experimental setup. The Raman tensors of the $\text{E}_\text{g}$ modes, $\text{E}^{a}_{g}$ and $\text{E}^{b}_{g}$, take the following form~\cite{Loudon1964}
\begin{equation}
R_{\text{E}^{a}_\text{g}} =
\begin{pmatrix}
  a & 0 \\ 
  0 & -a
\end{pmatrix}, 
\label{eq:R_1} 
\end{equation}
\quad
\begin{equation}
R_{\text{E}^{b}_\text{g}} =
\begin{pmatrix}
  0 & -a \\ 
  -a & 0
\end{pmatrix}.
\label{eq:R_2}
\end{equation}
Because of the symmetry group of the crystal it is possible to choose the coordinates $Q_{\text{E}^{a}_\text{g}}$ and $Q_{\text{E}^{b}_\text{g}}$ such that they transform, respectively, as $Q_{\text{E}^{b}_\text{u}}^2-Q_{\text{E}^{a}_\text{u}}^2$ and $Q_{\text{E}^{a}_\text{u}}Q_{\text{E}^{b}_\text{u}}$. With this choice, the phonon potential energy including anharmonic contributions reads
\begin{align} 
V &= \frac{\Omega_{\text{E}_\text{u}}^2}{2} (Q^{2}_{\text{E}^{a}_\text{u}} + Q^{2}_{\text{E}^{b}_\text{u}}) + \frac{\Omega_{\text{E}_\text{g}}^2}{2} (Q^{2}_{\text{E}^{a}_\text{g}} + Q^{2}_{\text{E}^{b}_\text{g}}) + cQ_{\text{E}^{a}_\text{g}}(Q_{\text{E}^{b}_\text{u}}^2-Q_{\text{E}^{a}_\text{u}}^2)+2cQ_{\text{E}^{b}_\text{g}}Q_{\text{E}^{a}_\text{u}}Q_{\text{E}^{b}_\text{u}} +\tilde{V}.
\label{eq:phononpotential}
\end{align}

Here, $\Omega_{\text{E}_\text{u}}$ and $\Omega_{\text{E}_\text{g}}$ are the eigenfrequencies of the IR-active $\text{E}_\text{u}$ and Raman-active $\text{E}_\text{g}$ modes, respectively, while the three-phonon coupling $c$ denotes the nonlinear phonon coupling of primary interest here. $\tilde{V}$ contains higher order anharmonicities and nonlinear phonon couplings at fourth order, as dictated by the $\bar{3}m$ point-group symmetry,
\begin{align}
\tilde{V}=&~ d_{a} Q^4_{\text{E}^{a}_\text{u}}+d_{b} Q^4_{\text{E}^{b}_\text{u}}+d_{c} Q^4_{\text{E}^{a}_\text{g}}+d_{d} Q^4_{\text{E}^{b}_\text{g}}\nonumber\\
& + d_{ab} Q^2_{\text{E}^{a}_\text{u}}Q^2_{\text{E}^{b}_\text{u}}+d_{ac} Q^2_{\text{E}^{a}_\text{g}}Q^2_{\text{E}^{a}_\text{u}}+d_{bc} Q^2_{\text{E}^{a}_\text{g}}Q^2_{\text{E}^{b}_\text{u}}+d_{ad} Q^2_{\text{E}^{b}_\text{g}}Q^2_{\text{E}^{a}_\text{u}}+d_{bd} Q^2_{\text{E}^{b}_\text{g}}Q^2_{\text{E}^{b}_\text{u}}\nonumber\\
&+d_{abc}Q^2_{\text{E}^{a}_\text{g}}Q_{\text{E}^{a}_\text{u}}Q_{\text{E}^{b}_\text{u}}+d_{abd}Q^2_{\text{E}^{b}_\text{g}}Q_{\text{E}^{a}_\text{u}}Q_{\text{E}^{b}_\text{u}}.
\end{align}
Ionic Raman scattering (IRS) is described by three-phonon coupling between two IR-active phonons and one Raman-active phonon, as in Eq.~\ref{eq:phononpotential}. The combined dynamics of the $\text{E}_\text{u}$ and $\text{E}_\text{g}$ modes for this process can be obtained by solving the coupled equations of motion resulting from Eq.~\ref{eq:general_equation_motion} with $R_{ij}=0$ and the phonon potential $V$ given by Eq.~\ref{eq:phononpotential}. The equations of motion then read
\begin{align}
\label{eq:Q_1_time_IRS}
 \ddot{Q}_{\text{E}^{a}_\text{u}} + \gamma_{\text{E}_\text{u}}\Dot{Q}_{\text{E}^{a}_\text{u}} + \Omega^2_{\text{E}_\text{u}}Q_{\text{E}^{a}_\text{u}} & = Z_{\text{E}^{a}_\text{u},x}{E}_{x}(t) + 2c Q_{\text{E}^{a}_\text{u}} Q_{\text{E}^{a}_\text{g}} - 2c Q_{\text{E}^{b}_\text{u}} Q_{\text{E}^{b}_\text{g}} - \partial_{Q_{\text{E}^{a}_\text{u}}} \tilde{V}, \\
\label{eq:Q_2_time_IRS}
 \ddot{Q}_{\text{E}^{b}_\text{u}} + \gamma_{\text{E}_\text{u}}\Dot{Q}_{\text{E}^{b}_\text{u}} + \Omega^2_{\text{E}_\text{u}}Q_{\text{E}^{b}_\text{u}} & = Z_{\text{E}^{b}_\text{u},y}E_{y}(t)  - 2c Q_{\text{E}^{b}_\text{u}} Q_{\text{E}^{a}_\text{g}} - 2c Q_{\text{E}^{a}_\text{u}} Q_{\text{E}^{b}_\text{g}} - \partial_{Q_{\text{E}^{b}_\text{u}}} \tilde{V},\\
\label{eq:Q_c_IRS}
 \ddot{Q}_{\text{E}^{a}_\text{g}} + \gamma_{\text{E}_\text{g}}\Dot{Q}_{\text{E}^{a}_\text{g}} + \Omega^2_{\text{E}_\text{g}}Q_{\text{E}^{a}_\text{g}} & = -c(Q_{\text{E}^{b}_\text{u}}^2-Q_{\text{E}^{a}_\text{u}}^2) - \partial_{Q_{\text{E}^{a}_\text{g}}} \tilde{V},\\
\label{eq:Q_d_IRS}
 \ddot{Q}_{\text{E}^{b}_\text{g}} + \gamma_{\text{E}_\text{g}}\Dot{Q}_{\text{E}^{b}_\text{g}} + \Omega^2_{\text{E}_\text{g}}Q_{\text{E}^{b}_\text{g}} & = - 2cQ_{\text{E}^{a}_\text{u}}Q_{\text{E}^{b}_\text{u}}-\partial_{Q_{\text{E}^{b}_\text{g}}} \tilde{V},
\end{align}
with $\gamma_{\text{E}_\text{u}}=0.2\times 2\pi$~THz$\cdot$rad and $\gamma_{\text{E}_\text{g}}=0.29\times 2\pi$~THz$\cdot$rad.
In contrast, for terahertz sum-frequency excitation (THz-SFE), we consider a purely harmonic phonon potential, as anharmonicities are disregarded in this process. Thus, only the harmonic terms from Eq.~\ref{eq:phononpotential} contribute to the equations of motion. Furthermore, in centrosymmetric crystals $\boldsymbol{Z}_\alpha=0$ for Raman-active phonons, and Eq.~\ref{eq:general_equation_motion} accordingly simplifies to
\begin{align}
\label{eq:Q_c_SFE}
 \ddot{Q}_{\text{E}^{a}_\text{g}} + \gamma_{\text{E}_\text{g}}\Dot{Q}_{\text{E}^{a}_\text{g}} + \Omega^2_{\text{E}_\text{g}}Q_{\text{E}^{a}_\text{g}} & =\epsilon_{0}a (E^2_{x}(t)-E^2_{y}(t)),\\
\label{eq:Q_d_SFE}
\ddot{Q}_{\text{E}^{b}_\text{g}} + \gamma_{\text{E}_\text{g}}\Dot{Q}_{\text{E}^{b}_\text{g}} + \Omega^2_{\text{E}_\text{g}}Q_{\text{E}^{b}_\text{g}} & = -2\epsilon_{0}aE_{x}(t)E_{y}(t).
\end{align}
The two processes differ in the driving force that the coherently excited $\text{E}_\text{u}$ modes produce for the Raman-active $\text{E}_\text{g}$ modes. In THz-SFE, the driving force is proportional to the square of the electric field (see Eqs.~\ref{eq:Q_c_SFE} and \ref{eq:Q_d_SFE}). In contrast, for IRS, the primary component of the driving force is proportional to the square of the IR-active phonon amplitude (see Eqs.~\ref{eq:Q_c_IRS} and \ref{eq:Q_d_IRS}). 

Finally, the mechanical angular momentum produced by the phonon modes can be obtained in terms of the phonon amplitudes~\cite{Juraschek2017}
\begin{equation}
\boldsymbol{L}(t) = \boldsymbol{Q} (t) \times \dot{\boldsymbol{Q}} (t),
\label{eq:phonon_angular_momentum}
\end{equation}
where the amplitude vectors are given by $\boldsymbol{Q}_{\text{E}_\text{u}}(t)=(Q_{\text{E}^{a}_\text{u}}(t),Q_{\text{E}^{b}_\text{u}}(t),0)$ and $\boldsymbol{Q}_{\text{E}_\text{g}}(t)=(Q_{\text{E}^{a}_\text{g}}(t),Q_{\text{E}^{b}_\text{g}}(t),0)$, respectively.



\subsubsection*{S3. Details of first-principle calculations}\label{sec:S3}
The general anharmonic phonon potential energy we use for fitting reads
\begin{align} 
V =&~\frac{\Omega_{\text{E}_\text{u}}^2}{2} (Q^{2}_{\text{E}^{a}_\text{u}} + Q^{2}_{\text{E}^{b}_\text{u}}) + \frac{\Omega_{\text{E}_\text{g}}^2}{2} (Q^{2}_{\text{E}^{a}_\text{g}} + Q^{2}_{\text{E}^{b}_\text{g}})\nonumber\\
&+d_{a} Q^4_{\text{E}^{a}_\text{u}}+d_{b} Q^4_{\text{E}^{b}_\text{u}}+d_{c} Q^4_{\text{E}^{a}_\text{g}}+d_{d} Q^4_{\text{E}^{b}_\text{g}}\nonumber\\
&+c_{abc}Q_{\text{E}^{a}_\text{g}}Q_{\text{E}^{a}_\text{u}}Q_{\text{E}^{b}_\text{u}}+c_{abd}Q_{\text{E}^{b}_\text{g}}Q_{\text{E}^{a}_\text{u}}Q_{\text{E}^{b}_\text{u}}\nonumber\\
&+c_{ac}Q_{\text{E}^{a}_\text{g}}Q^2_{\text{E}^{a}_\text{u}}+c_{ad}Q_{\text{E}^{b}_\text{g}}Q^2_{\text{E}^{a}_\text{u}}+c_{bc}Q_{\text{E}^{a}_\text{g}}Q^2_{\text{E}^{b}_\text{u}}+c_{bd}Q_{\text{E}^{b}_\text{g}}Q^2_{\text{E}^{b}_\text{u}}\nonumber\\
&+d_{ab} Q^2_{\text{E}^{a}_\text{u}}Q^2_{\text{E}^{b}_\text{u}} + d_{ac} Q^2_{\text{E}^{a}_\text{g}}Q^2_{\text{E}^{a}_\text{u}}+d_{ad} Q^2_{\text{E}^{b}_\text{g}}Q^2_{\text{E}^{a}_\text{u}}+d_{bc} Q^2_{\text{E}^{a}_\text{g}}Q^2_{\text{E}^{b}_\text{u}}+d_{bd} Q^2_{\text{E}^{b}_\text{g}}Q^2_{\text{E}^{b}_\text{u}}\nonumber\\
&+d_{abc}Q^2_{\text{E}^{a}_\text{g}}Q_{\text{E}^{a}_\text{u}}Q_{\text{E}^{b}_\text{u}}+d_{abd}Q^2_{\text{E}^{b}_\text{g}}Q_{\text{E}^{a}_\text{u}}Q_{\text{E}^{b}_\text{u}}.
\label{eq:nonrotatedphononpotential}
\end{align}
This general potential with mixed components of the doubly degenerate modes can be brought back into the form of Eq.~\ref{eq:phononpotential} with a coordinate transformation. 
Next, we compute the total energy on a $11\times11\times11$ grid defined by atomic displacements along the eigenvectors of the rotated $\text{E}_\text{u}$ and $\text{E}_\text{g}$ modes. The resulting anharmonic potential explicitly includes coupling terms between the doubly degenerate $\text{E}_\text{u}$ modes and each of the $\text{E}_\text{g}$ modes separately, while no coupling is considered between the individual $\text{E}_\text{g}$ modes themselves, which is deemed to be small.

\begin{table} 
	\centering
	\caption{\textbf{Summary of calculated phonon properties and anharmonic coefficients.}
		Calculated eigenfrequencies in THz, single-mode anharmonicities and nonlinear phonon couplings in meV/(\AA{}$\sqrt{u})^n$, $n$ being the order of the phonon amplitude and $u$ the atomic mass unit, and mode effective charges in $e/\sqrt{u}$, where $e$ is the elementary charge.}
	\label{table} 
\resizebox{\textwidth}{!}{
\begin{tabular}{lllllllllllllllllll}
\hline
 & $c_{abc}$ & $c_{abd}$ & $c_{ac}$ & $c_{ad}$ & $c_{bc}$& $c_{bd}$ & $d_{a}$ & $d_{b}$ & $d_{c}$ & $d_{d}$ & $d_{ab}$ & $d_{ac}$ &  $d_{ad}$ & $d_{bc}$ & $d_{bd}$ & $d_{abc}$ &  $d_{abd}$ & $Z$ \\ 
\hline
$\text{E}^{a}_\text{u}(2.6)$ & & & & & & & 0.3 &&&& 0.5 &&&&&& &1.04\\
$\text{E}^{b}_\text{u}(2.6)$ & & & & & & & & 0.3 &&& 0.5 &&&&&&& 1.04\\
$\text{E}^{a}_\text{g}(4.2)$ &  -1.1 & & 1.3 & & -1.3 & &&& -0.02 &&& -0.02 && 0.2&&0.07 & & 0\\
$\text{E}^{b}_\text{g}(4.2)$ & & -2.6 &  & -0.6 & & 0.6 &&&& -0.02 &&&0.2&  &-0.02&  & -0.07& 0\\
\hline
\end{tabular}
}\label{ext:table}
\end{table}

We perform a three-step fitting procedure on the resulting potential energy landscape to extract single-mode anharmonicities and nonlinear phonon couplings from the phonon potential energy $V$ in Eq.~\ref{eq:nonrotatedphononpotential}. First, we obtain the single-mode anharmonicities for the $\text{E}_\text{u}$ and $\text{E}_\text{g}$ modes by fitting the potential energy landscape for each mode individually. We use the total energy from the grid points where displacements occur along a single-phonon mode, with the others fixed to zero. From this, we obtain the coefficients $\Omega_{\text{E}_\text{u}}, \Omega_{\text{E}_\text{g}}, d_{a}, d_{b}, d_{c}, d_{d}$. We include the phonon frequencies in the fitting to ensure that they match those computed with \textsc{phonopy}. Next, we focus on the term that involved only the $\text{E}_\text{u}$ modes, $Q_{\text{E}^{a}_\text{u}}^2 Q_{\text{E}^{b}_\text{u}}^2$, by fitting the potential energy landscape to the total energy values where only the $\text{E}_\text{u}$ modes are displaced, with the $\text{E}_\text{g}$ modes held at zero. The single-mode anharmonicities of the first step are kept fixed, allowing us to extract the coefficient $d_{ab}$. Finally, we fit the potential energy landscape to the terms in $V$ (in Eq.~\ref{eq:nonrotatedphononpotential}) that involve the three phonon modes for each of the $\text{E}_\text{g}$ modes, using total energy values from the grid points where all modes are displaced. With the coefficients from the previous steps held constant, we determine the nonlinear phonon couplings $d_{ac}$, $d_{ad}$, $d_{bc}$, $d_{bd}$,$d_{abc}$, $d_{abd}$. Table~S1 in Supplementary Information summarizes the computed phonon properties and anharmonic coefficients. 

We note that the calculated phonon frequencies are off by about 30~\%{} and 5\%{} for the $\text{E}_\text{u}$ and $\text{E}_\text{g}$ modes, respectively, with respect to experimental values. This discrepancy is known to arise in DFT calculations for this class of material \cite{Wang2012, Boulares2018}. Accordingly, also the nonlinear couplings could be varying by two-digit percentages compared to experiment. Such a variation would lead to small quantitative changes in the calculated angular momentum, but would, however, leave the analysis and interpretation of the mechanisms unchanged. Additionally, as pointed out in \cite{Wang2012}, temperature can influence phonon eigenfrequencies. Therefore, part of the discrepancy between the DFT and experimental results likely arises from the fact that the DFT calculations are performed at zero temperature, whereas the experiments are conducted at room temperature.

\subsubsection*{S4. Analytical model for THz transmission measurement of $\text{Bi}_2\text{Se}_3$} \label{sec:S4}
To analyze the THz transmittance of $\text{Bi}_2\text{Se}_3$ (Extended Data Fig.~1d) and determine the phonon lifetime of the $\text{E}_\text{u}$ mode $\tau_{\text{IR}}$, we derive the analytical expression based on Fresnel coefficients. The analytical solution accounts for all internal reflections within both the $\text{Bi}_2\text{Se}_3$ sample and the sapphire substrate. Owing to the topological insulating properties of $\text{Bi}_2\text{Se}_3$, the material is represented as a three-layer system comprising an insulating bulk sandwiched between two conductive topological surface states (TSS) \cite{Chen2022a, Pogna2021}. Figure~S4 illustrates the layered structure and defines the associated transmission and reflection coefficients. The multilayer stack consists of a TSS ($n_1$, refractive index), a $\text{Bi}_2\text{Se}_3$ bulk layer ($n_2$, refractive index), a TSS ($n_1$, refractive index), and a sapphire substrate ($n_3$, refractive index), surrounded by air ($n_0$, refractive index) from both sides. The THz pulse propagates at normal incidence through the layers in the sequence $n_0\rightarrow n_1\rightarrow n_2 \rightarrow n_1\rightarrow n_3\rightarrow n_0$.

The reference transmission function for sapphire, $T_{\text{sapph}}(\omega)$, is derived for a single layer, accounting for multiple beam interference within the substrate
\begin{equation}
    T_{\text{sapph}}(\omega)= \frac{t_{03}t_{30}e^{i\varphi_3}}{1-r^2_{30}e^{i2\varphi_3}}.
\end{equation}
Here, $t_{ij}(\omega)$ and $r_{ij}(\omega)$ denote the Fresnel transmission and reflection coefficients for a wave propagating from media $i$ to $j$, defined as $t_{ij}(\omega)=2n_i/(n_i+n_j)$ and $r_{ij}(\omega)=(n_i-n_j)/(n_i+n_j)$. The phase $\varphi_3$ accounts for the THz pulse propagation through the sapphire layer of thickness $d_3$ = 500~$\mu$m, given by $\varphi_3(\omega)=(\omega n_3(\omega) d_3)/c_\text{o}$, where $c_\text{o}$ is the speed of light in vacuum.

To derive the sample transmission function $T_{\text{total}}(\omega)$, we employ an iterative approach that systematically reduces a multilayer system to an effective single-layer model by incorporating the contribution of other layers into the modified transmission and reflection coefficients at the interfaces. As the first step, the $\text{Bi}_2\text{Se}_3$ sample is treated as an effective interface between air and the sapphire substrate, characterized by the transmission ($T_{03}$) and reflection ($R_{30}$) coefficients. This simplification reduces the system to an equivalent single-layer sapphire ($n_3$) model, analogous to the reference case, so the THz transmission function through the sample and the substrate $T_{\text{total}}(\omega)$ can be expressed as
\begin{equation}
    T_{\text{total}}(\omega)= \frac{T_{03}t_{30}e^{i\varphi_3}}{1-r_{30}R_{30}e^{i2\varphi_3}}.
\end{equation}

\begin{figure} 
	\centering
	\includegraphics[width=\textwidth]{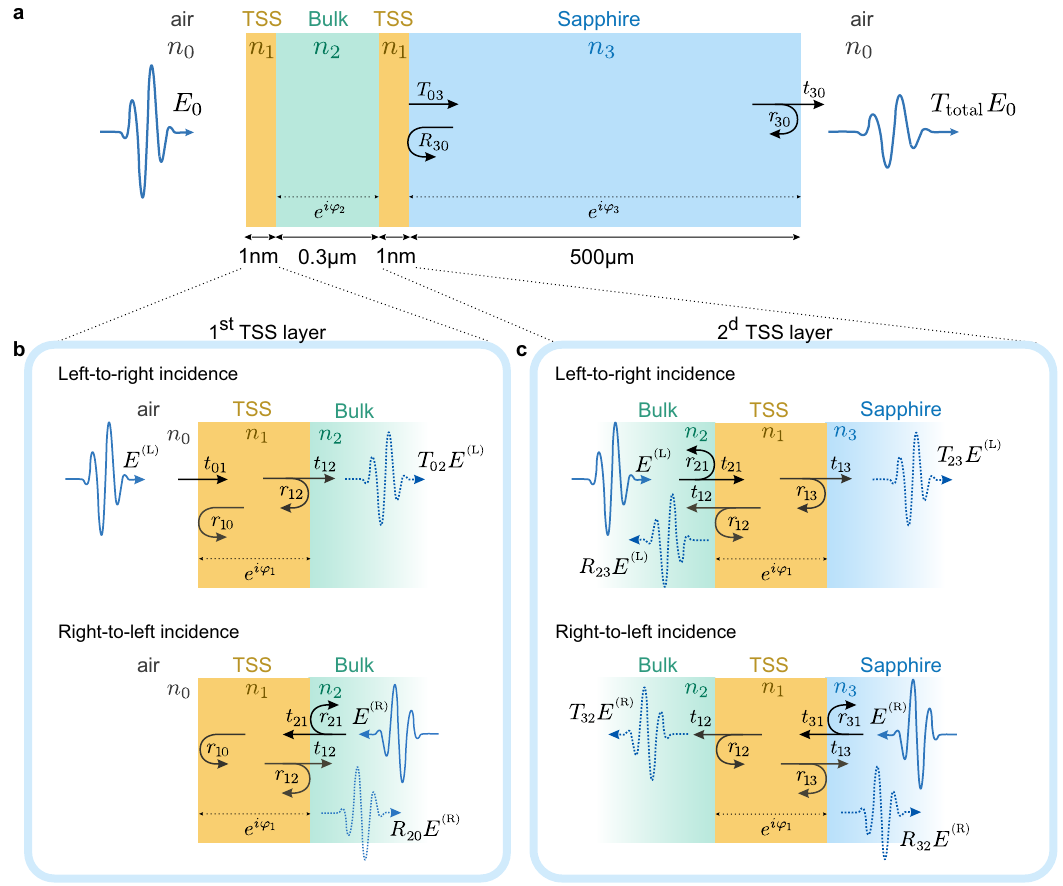} 

	\caption{\textbf{Multilayer thin-film interference model.}
		\textbf{a}, Schematic of multilayer system of $\textrm{Bi}_2\textrm{Se}_3$ sample on the sapphire substrate and the transmitted THz field (blue waveform). $\textrm{Bi}_2\textrm{Se}_3$ is represented as a three-layer system of an insulating bulk layer sandwiched between two conductive surface layers (TSS). \textbf{b}-\textbf{c}, Transmission and reflection of the THz pulse from the first and second TSS layers, respectively. $t_{ij}$ and $r_{ij}$ are the Fresnel transmission and reflection coefficients for a wave propagating from medium $i$ into medium $j$. $T_{ij}$ and $R_{ij}$ accounts for all internal reflections between the media $i$ and $j$.}
	\label{fig:S4} 
\end{figure}

To determine $T_{03}$ and $R_{30}$, we separately consider a three-layer $\text{Bi}_2\text{Se}_3$ sample and reduce its complexity to a single layer-model of the bulk ($n_2$) sandwiched between air ($n_0$)  and sapphire ($n_3$), with effective interfaces defined by the TSSs (Fig.~S4a). Here we introduce the transmission and reflection coefficients at the air-bulk ($T_{02}$, $R_{20}$), and bulk-sapphire ($T_{23}$, $R_{23}$, $T_{32}$, $R_{32}$) interfaces, so the transmission and reflection of the $\text{Bi}_2\text{Se}_3$ sample can be written as: 
\begin{equation}
    T_{03}(\omega)= \frac{T_{02}T_{23}e^{i\varphi_2}}{1-R_{20}R_{23}e^{i2\varphi_2}}, \quad R_{30}(\omega)= R_{32}+\frac{T_{32}T_{23}R_{20}e^{i2\varphi_2}}{1-R_{20}R_{23}e^{i2\varphi_2}},
\label{eq:joinTR}
\end{equation}
where $\varphi_2(\omega)=(\omega n_2(\omega) d_2)/c_\text{o}$ stands for the phase accumulated by the THz pulse propagating through the $\text{Bi}_2\text{Se}_3$ bulk with a thickness of $d_2$ = 0.3~$\mu$m.
The frequency-dependent coefficients  $T_{02}(\omega)$, $R_{20}(\omega)$ and $T_{23}(\omega)$, $R_{23}(\omega)$, $T_{32}(\omega)$, $R_{32}(\omega)$ are explicitly derived by considering two TSS layers independently: one interfacing with air and the bulk, and the other with the bulk and sapphire (see Fig.~S4b,c). These coefficients can be expressed in term of the Fresnel coefficients as follows
\begin{equation}
    T_{02}(\omega)= \frac{t_{01}t_{12}e^{i\varphi_1}}{1-r_{10}r_{12}e^{i2\varphi_1}}, \quad R_{20}(\omega)= r_{21}+\frac{t_{21}t_{12}r_{10}e^{i2\varphi_1}}{1-r_{10}r_{12}e^{i2\varphi_1}},
\end{equation}
\begin{equation}
    T_{23}(\omega)= \frac{t_{21}t_{13}e^{i\varphi_1}}{1-r_{13}r_{12}e^{i2\varphi_1}}, \quad R_{23}(\omega)= r_{21}+\frac{t_{21}t_{12}r_{13}e^{i2\varphi_1}}{1-r_{13}r_{12}e^{i2\varphi_1}},
\end{equation}
\begin{equation}
    T_{32}(\omega)= \frac{t_{31}t_{12}e^{i\varphi_1}}{1-r_{13}r_{12}e^{i2\varphi_1}}, \quad R_{32}(\omega)= r_{31}+\frac{t_{31}t_{13}r_{12}e^{i2\varphi_1}}{1-r_{13}r_{12}e^{i2\varphi_1}},
\end{equation}
where $\varphi_1(\omega)=(\omega n_1(\omega) d_1)/c_\text{o}$ with $d_1$ = 1~nm.

Therefore, using this iterative approach, we reduce the model’s complexity and derive the THz transmission function $T_{\text{total}}(\omega)$ in terms of only Fresnel coefficients and phase factors. The calculated THz transmission function of $\text{Bi}_2\text{Se}_3$, presented in Extended Data Fig.~1d, is determined by 
\begin{equation}
    T_{\text{Bi}_2\text{Se}_3}(\omega)= T_{\text{total}}(\omega)/T_{\text{sapph}}(\omega).
\end{equation}

To model the $\text{Bi}_2\text{Se}_3$ measured transmittance obtained from the windowed time-domain EOS traces (see Extended Data Fig.~1f), we keep only the contribution from the fundamental transmitted THz pulse by retaining only the numerators in $T_{\text{total}}(\omega)$ and $T_{\text{sapph}}(\omega)$. The sample and substrate transmission functions therefore take the form
\begin{equation}
    T_{\text{total}}(\omega)= T_{03}t_{30}e^{i\varphi_3},
\end{equation}
\begin{equation}
    T_{\text{sapph}}(\omega)= t_{03}t_{30}e^{i\varphi_3},
\end{equation}
respectively, leading to the $\text{Bi}_2\text{Se}_3$ transmission function $T_{\text{Bi}_2\text{Se}_3}(\omega)=T_{03}(\omega)/t_{03}(\omega)$.

The dielectric functions for the distinct layers of the $\text{Bi}_2\text{Se}_3$ sample are defined using different models tailored to their properties. The Lorentz oscillator model is employed for the insulating $\text{Bi}_2\text{Se}_3$ bulk $\varepsilon_\text{bulk}(\omega)$, while the Drude model is used for the conducting topological surface states $\varepsilon_\text{TSS}(\omega)$. The dielectric functions for air and sapphire are set as $\varepsilon_\text{air}(\omega)$ = 1 and $\varepsilon_\text{sapph}(\omega)$ = 3.
For the topological surface states, the dielectric function is given by
\begin{equation}
\varepsilon_\text{TSS}(\omega)=1-  \frac{\omega^2_{\text{pl}}}{\omega^2+i\gamma\omega},
\end{equation}
with $\omega_{\text{pl}}$=1050~THz and $\gamma$ = 20~THz. The introduction of conducting TSS layers at the interfaces allows to capture the reduction in the THz transmission across the spectrum.
For the $\text{Bi}_2\text{Se}_3$ bulk, the dielectric function is set as
\begin{equation}
\varepsilon_\text{bulk}(\omega)=\varepsilon_{\infty}+  \frac{S\omega^2_{\text{IR}}}{\omega^2_{\text{IR}}-\omega^2-i\gamma_{\text{IR}}\omega},
\end{equation}
with $\varepsilon_{\infty}$ = 40, $S$ = 100.5, $\omega_{\text{IR}}$ = 1.97~THz and $\gamma_{\text{IR}}$ = 0.20~THz. The parameters $\omega_{\text{IR}}$, $\gamma_{\text{IR}}$, S are the frequency of IR-active $\text{E}_\text{u}$ phonon, its damping rate and the oscillator strength, respectively. These parameters of the dielectric functions were adjusted to model the experimental data, focusing on determining the phonon frequency $\omega_{\text{IR}}$ and phonon damping rate $\gamma_{\text{IR}}$. The $\text{E}_\text{u}$ phonon lifetime is determined as $\tau_{\text{IR}}=1/(2\pi\gamma_{\text{IR}})$ and equals 0.79~ps.

\newpage

\begin{figure} 
	\centering
	\includegraphics[width=0.6\textwidth]{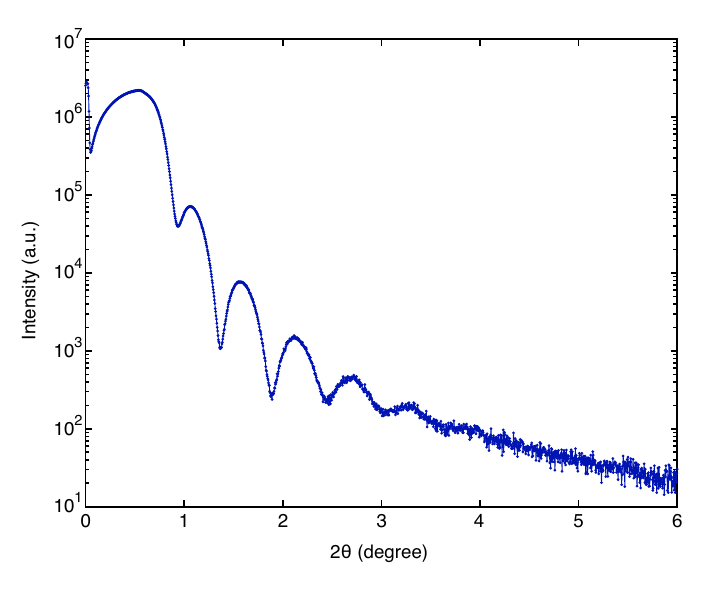} 

	\caption{\textbf{X-ray reflectivity (XRR).}
		XRR curve of the used $\textrm{Bi}_2\textrm{Se}_3$ sample. From the periodicity of the oscillations, a $\textrm{Bi}_2\textrm{Se}_3$ thickness $d=$15.0~nm is determined.}
	\label{fig:S5} 
\end{figure}

\begin{figure} 
	\centering
	\includegraphics[width=0.6\textwidth]{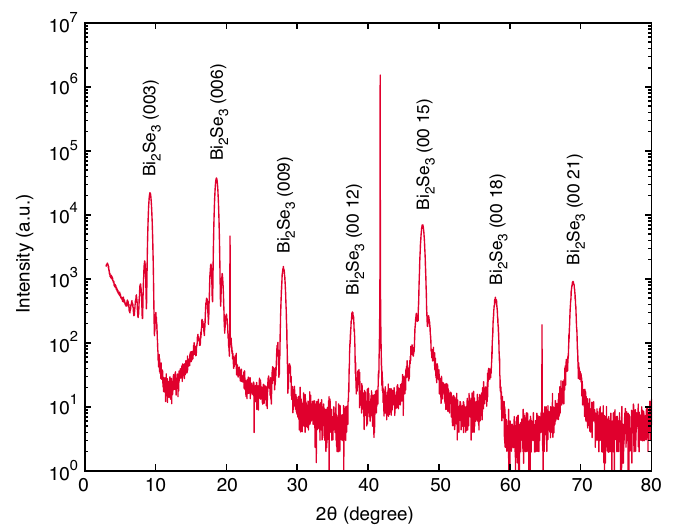} 

	\caption{\textbf{X-ray Diffraction (XRD).}
		Symmetric $2\theta/\theta$ XRD curve of the investigated sample with numerous peaks from the epilayer, evidencing the correct $\textrm{Bi}_2\textrm{Se}_3$ stoichiometry as well as the single-crystal nature of the $\textrm{Bi}_2\textrm{Se}_3$ film.}
	\label{fig:S6} 
\end{figure}

\begin{figure} 
	\centering
	\includegraphics[width=\textwidth]{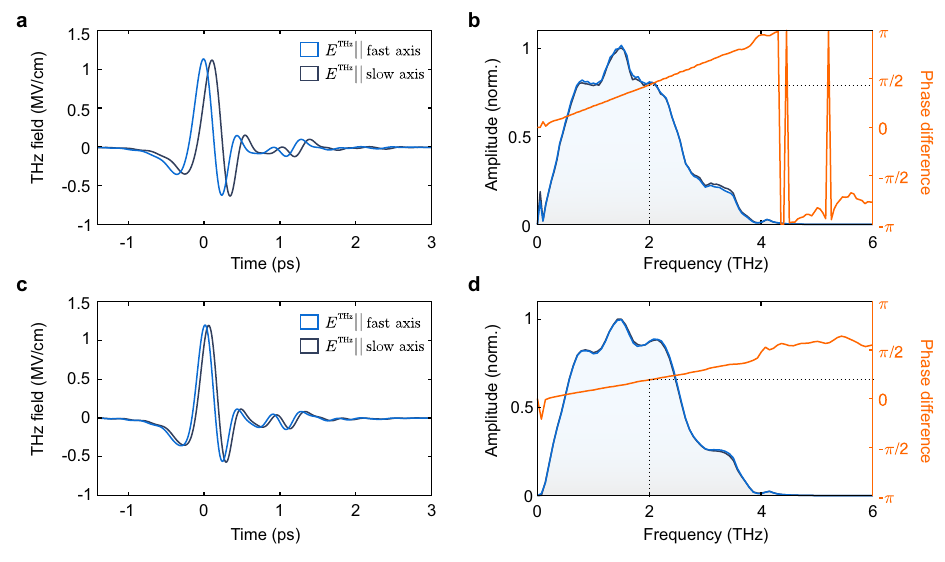} 

	\caption{\textbf{Characterization of y-cut quartz waveplates by THz electro-optic sampling.}
		\textbf{a}, Linearly-polarized THz electric fields after propagating through a 700~$\mu$m-thick y-cut quartz waveplate, measured via electro-optic sampling in a 50~$\mu$m-thick z-cut quartz detection crystal. Blue and black traces correspond to THz fields aligned with the fast and slow axes of y-cut quartz, respectively. \textbf{b}, Corresponding Fourier transforms of the THz fields and their relative phase difference (orange). At 2~THz, the phase difference is about $\pi/2$, indicating the generation of a circularly polarized THz pulse when the incident field is oriented at $45^\circ$ to the fast and slow axes. \textbf{c}-\textbf{d}, Same measurements as \textbf{a}–\textbf{b}, but for a 380~$\mu$m-thick y-cut quartz waveplate. The phase difference at 2~THz results in the generation of an arbitrarily elliptical THz polarization state when the incident THz field is oriented at $45^\circ$ relative to the crystal axes.}
	\label{fig:S7} 
\end{figure}

\begin{figure} 
	\centering
	\includegraphics[width=\textwidth]{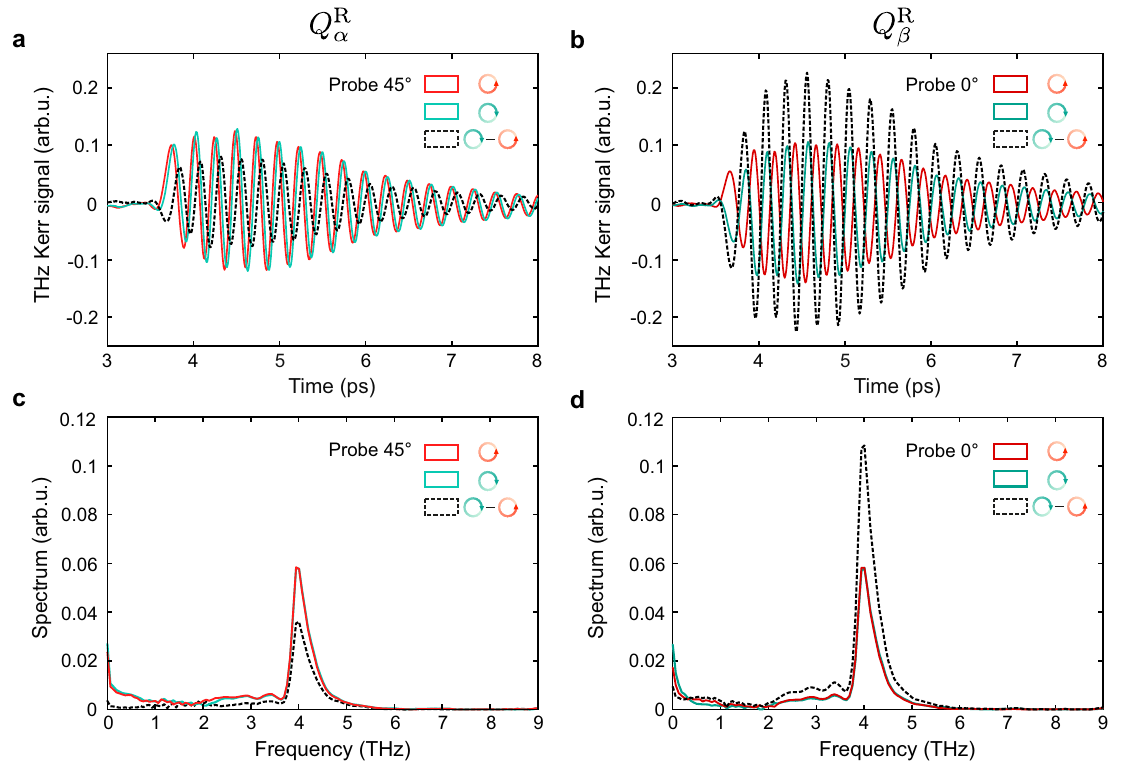}
	\caption{\textbf{Absence of transient magnetic contribution in the Kerr response.}
		THz Kerr signals measured under right- (red) and left- (blue) circularly  polarized THz excitation for probe polarizations of 45$^\circ$ and 0$^\circ$ (data presented in Extended Data Fig.4f,i) and their difference (dashed black) in the time (panels \textbf{a},\textbf{b}) and frequency (panels \textbf{c},\textbf{d}) domains. A pump-probe time drift in the RCP and LCP measurements introduces a small temporal shift between nearly identical Kerr traces. For  45$^\circ$ probe polarization, RCP and LCP THz pulses drive the $Q^{\text{R}}_{\alpha}$ component with the same phase, but the temporal shift causes the subtraction to produce a residual oscillation of the same shape. For  0$^\circ$ probe, the opposite phases of the driven $Q^{\text{R}}_{\beta}$ component lead the subtraction to enhance the oscillation. In both cases, the RCP–LCP difference shows no helicity-dependent slowly-varying background, confirming the absence of any magnetically induced Kerr contribution on the order of the phonon coherence time.}
	\label{fig:S8} 
\end{figure}


\clearpage 

\paragraph{Caption for Movie S1.} \label{mov:S1}
\textbf{Experimental data: THz electric field and measured $\text{E}_\text{g}$ phonon trajectories}
Animated version of Figure~1e,f. Trajectory of the THz excitation pulse’s electric field vector (left; Fig.~1e), measured via polarization-resolved electro-optic sampling, and the corresponding $\text{E}_\text{g}$ phonon trajectory (right; Fig.~1f), measured via the THz-induced Kerr effect measurement. A shared time reference was obtained from modeling the phononic excitation process (Fig.~3b).

\paragraph{Caption for Movie S2.} \label{mov:S2}
\textbf{Simulation: Coupled $\text{E}_\text{u}$ phonon and $\text{E}_\text{g}$ phonon trajectories.}
Animated version of Extended Data Fig.~6b,d. $\text{E}_\text{u}$ phonon trajectory (left; Extended Data Fig.~6b) calculated from the experimental RCP THz excitation field (Fig.~1e) and the corresponding $\text{E}_\text{g}$ phonon trajectory (right; Extended Data Fig.~6d) driven through the lowest-order anharmonic lattice potential $V(\boldsymbol{Q}^{\text{IR}},\boldsymbol{Q}^{\text{R}}) = c[(Q^{\text{IR}}_y)^2-(Q^{\text{IR}}_x)^2]Q^{\text{R}}_{\alpha}+2c\,Q^{\text{IR}}_x Q^{\text{IR}}_y Q^{\text{R}}_{\beta}$.

\paragraph{Caption for Movie S3.} \label{mov:S3}
\textbf{Time evolution of phonon helicity reversal based on ab-initio DFT.}
\textbf{a}-\textbf{b}, The animation displays the time evolution of the atomic displacements for the $\text{E}_\text{u}$ (\textbf{a}) and $\text{E}_\text{g}$ (\textbf{b}) phonon modes in $\text{Bi}_2\text{Se}_3$ unit cell. These dynamics were computed by solving the coupled equations of motion (Eqs.~\ref{eq:Q_1_time_IRS}--\ref{eq:Q_d_IRS}) with the full anharmonic potential $V$ (Eq.~\ref{eq:nonrotatedphononpotential}), projected onto eigenvectors derived from \textit{ab-initio} DFT (Fig.~S2c,d). Phonon amplitudes are independently normalized for clearer visibility. The visualization demonstrates the counter-rotating nature of the interaction: the driven $\text{E}_\text{u}$ mode follows the handedness of the driving field (Fig.~2a, red trace), while the induced $\text{E}_\text{g}$ mode rotates in the opposite direction. \textbf{c}-\textbf{d}, Corresponding quantitative angular momentum dynamics for the $\text{E}_\text{u}$ (\textbf{c}) and $\text{E}_\text{g}$ (\textbf{d}) modes (presented in Fig.3c).



\end{document}